\documentclass[fleqn,usenatbib]{mnras}

\usepackage{newtxtext,newtxmath}

\usepackage[T1]{fontenc}

\DeclareRobustCommand{\VAN}[3]{#2}
\let\VANthebibliography\thebibliography
\def\thebibliography{\DeclareRobustCommand{\VAN}[3]{##3}\VANthebibliography}

\usepackage{graphicx}	
\usepackage{amsmath}	
\usepackage[normalem]{ulem}
\usepackage{lineno}
\usepackage{orcidlink}
\usepackage{adjustbox}
\usepackage{natbib}

\def\planck{\textit{Planck}}
\newcommand{\sqdeg}{\ensuremath{{\rm deg^2}}}
\providecommand{\sorthelp}[1]{}

\title[Detection of polarization in stacked PGCC clumps]{Polarization fraction of \textit{Planck} Galactic cold clumps  and forecasts for the Simons Observatory}

\author[J. Clancy et al.]{
J. Clancy~\orcidlink{0000-0002-9711-9969},$^{1}$\thanks{E-mail: clancyj1@student.unimelb.edu.au}
G. Puglisi~\orcidlink{0000-0002-0689-4290},$^{2}$
S. E. Clark~\orcidlink{0000-0002-7633-3376},$^{3,4}$
G. Coppi~\orcidlink{0000-0002-6362-6524},$^{5,6}$
G. Fabbian~\orcidlink{0000-0002-3255-4695},$^{7,8}$
C. Herv\'ias-Caimapo~\orcidlink{0000-0002-4765-3426},${^{9}}$
\newauthor{}
J. C. Hill~\orcidlink{0000-0002-9539-0835},$^{10}$
F. Nati~\orcidlink{0000-0002-8307-5088},$^{5}$
C. L. Reichardt~\orcidlink{0000-0003-2226-9169}$^{1}$
\\
$^{1}$School of Physics, The University of Melbourne, Parkville VIC 3030, Australia\\
$^{2}$Dipartimento di Fisica e Astronomia, Universit\'a degli Studi di Catania, Via S. Sofia,64, 95123, Catania, Italy\\
$^{3}$Department of Physics, Stanford University, Stanford, CA 94305, USA\\
$^{4}$Kavli Institute for Particle Astrophysics \& Cosmology (KIPAC), Stanford University, Stanford, CA 94305, USA\\
$^{5}$Department of Physics, University of Milano-Bicocca, Piazza della Scienza 3, 20126 Milano, Italy\\
$^{6}$National Institute for Nuclear Physics (INFN), Sezione di Milano-Bicocca, Piazza della Scienza 3, 20126 Milano, Italy\\
$^{7}$Center for Computational Astrophysics, Flatiron Institute, 162 5th Avenue, 10010, New York, NY, USA\\
$^{8}$School of Physics and Astronomy, Cardiff University, The Parade, Cardiff, CF24 3AA, UK\\
$^{9}$Instituto de Astrof\'isica and Centro de Astro-Ingenier\'ia, Facultad de F\'isica, Pontifica Universidad Cat\'olica de Chile, Av. Vicu\~na Mackenna 4860, \\ ~\,7820436 Macul, Santiago, Chile\\
$^{10}$Department of Physics, Columbia University, New York, NY, USA 10027\\
}

\date{Accepted 2023 July 6. Received 2023 June 5; in original form 2023 March 4}

\pubyear{2023}

\begin{document}
\label{firstpage}
\pagerange{\pageref{firstpage}--\pageref{lastpage}}
\maketitle

\defcitealias{ade_2019}{SO19}
\defcitealias{PlanckXXXVIII2016}{Planck Collaboration XXXVIII 2016}
\defcitealias{Planck2015XXVIII}{Planck Collaboration XXVIII 2015}
\defcitealias{Planck2015X}{Planck Collaboration X 2015}
\defcitealias{Planck2018III}{Planck Collaboration III 2018}
\defcitealias{Planck2018IV}{Planck Collaboration IV 2018}
\defcitealias{Planck2018XI}{Planck Collaboration XI 2018}
\defcitealias{Planck2018XII}{Planck Collaboration XII 2018}
\defcitealias{Planck2020I}{Planck Collaboration I 2020}
\defcitealias{Planck2020Npipe}{Planck Collaboration LVII 2020}
\defcitealias{PlanckIntXLVIII2016}{Planck Collaboration Int. Results XLVIII 2016}

\begin{abstract}
We measure the mean-squared polarization fraction of a sample of 6282 Galactic cold clumps at 353$\,$GHz, consisting of \textit{Planck} Galactic cold clump (PGCC) catalogue category 1 objects (flux densities measured with signal-to-noise ratio (S/N) $>4$). 
At 353$\,$GHz we find the mean-squared polarization fraction, which we define as the mean-squared polarization divided by the mean-squared intensity, to be $[4.79\pm0.44]\times10^{-4}$ equation to an $11\,\sigma$ detection of polarization.
We test if the polarization fraction depends on the clumps' physical properties, including flux density, luminosity, Galactic latitude and physical distance. 
We see a trend towards increasing polarization fraction with increasing Galactic latitude, but find no evidence that polarization depends on the other tested properties. The Simons Observatory, with angular resolution of order 1$^\prime$ and noise levels between 22 and $54\,\mu$K$-{\textrm{arcmin}}$ at high frequencies, will substantially enhance our ability to determine the magnetic field structure in Galactic cold clumps. 
At $\ge5\,\sigma$ significance, we predict the Simons Observatory will detect at least $\sim$12,000 cold clumps in intensity and $\sim$430 cold clumps in polarization. 
This number of polarization detections would represent a two orders of magnitude increase over the current \planck{} results. 
We also release software that can be used to mask these Galactic cold clumps in other analyses.

\end{abstract}

\begin{keywords}
polarization -- ISM: magnetic fields -- (Galaxy:) solar neighbourhood -- stars: protostars -- cosmology: observations
\end{keywords}



\section{Introduction}\label{sec:intro}

In current models of star formation, stars form within cold, dense clumps in filamentary and clumpy molecular clouds \citep{Andre2010,Hacar2022}. 
The large-scale magnetic fields pervading these clumps are expected to affect the movement of gas and subsequently the star formation processes e.g., providing directionality of gas flows \citep{SeifriedandWalch2015}, providing pressure support against gravitational instability \citep{NakanoandNakamura1978} and changing the characteristics of shocks \citep{Inoue2009}. Observations of the magnetic fields in these cold clumps are crucial to furthering our understanding of the role magnetic fields play in the evolution of these clumps and star formation. 

The magnetic field structure can be inferred by looking at the polarized patterns in the clumps' thermal emission. The thermal emission of dust grains in these clumps is partially polarized due to the asymmetric grains rotational axis being aligned perpendicular to the magnetic field lines, allowing the magnetic field structure to be probed by mapping the polarization field at mm/sub-mm wavelengths \citep{caselli2011}.
By combining these polarization measurements of the magnetic field with other observations (such as the position of dense protostars), observations of the polarized dust offer the opportunity to begin understanding the interplay between magnetic fields and the evolution of cold clumps and their host filaments and clouds \citep{ward-thompson1994}.

Large area cosmic microwave background (CMB) surveys with observation bands extending towards the sub-mm \citep[$\nu\gtrsim250\,$GHz;][]{Choi2015} are well-suited to making population-level studies of the thermal dust emission from Galactic cold clumps. The most complete data at present comes from the \textit{Planck} satellite \citep{planck2014-a12, planck2016-XLVIII, planck2016-l11A}, with full-sky polarization maps at 353$\,$GHz included in the fourth product release \citep[PR4;][]{planck2020-LVII}. \citet{planck2014-a37} identified 13,242 Galactic cold clumps (PGCCs) with low dust temperatures (6--20$\,$K) and moderately dense column densities when compared to other kinds of star forming clouds \citep{andre2014}. These clumps are prime candidates for probing the formation and evolution of protostellar cores.

A number of higher resolution, targeted observations have been made to examine the internal structure of individual PGCCs with the \textit{Herschel} satellite \citep{Juvela2012, Juvela2015, Juvela2018a,Montillaud2015,Rivera-Ingraham2016} and the SCUBA-II instrument at the James Clerk Maxwell Telescope \citep{Liu2018, Juvela2018b}. These observations are important as the \textit{Planck} data, with 5$\,$arcmin resolution, only resolve the magnetic field substructure for a limited number of very nearby cold clumps. Subsequent literature has explored properties from chemical composition \citep{Wakelam2021} to column density and temperature mapping for signs of star formation \citep{Zahorecz2016}. However while these studies, with individual or small numbers of clumps, have greatly expanded our knowledge of these objects, it is also crucial to study the population-level statistics for the polarization and magnetic field structure within star-forming clouds.

We present in this work the first detection of a population-average mean-squared polarization fraction, $\langle P^2\rangle/\langle I^2\rangle$, of a sample of \textit{Planck} Galactic cold clumps. To achieve this, we utilise a stacking technique to measure the mean-squared intensity and polarization signal arising from these clumps within \textit{Planck} total intensity and polarization maps. We follow this with an exploration of whether the polarization of PGCCs depends on properties including flux density, luminosity, Galactic latitude and physical distance. Throughout these tests we also discuss where the Simons Observatory (SO) will offer significant improvements.

This paper is organised as follows. In section~\ref{sec:dataset}, we describe the PGCC catalogue and the selection criteria we use to form our sample, as well as the 353$\,$GHz \textit{Planck} maps we use for our measurements and an introduction of the SO. We discuss the stacking procedure in section~\ref{sec:Methods}, along with the procedure for background correction and error estimation. We present our results in section~\ref{sec:results} and forecast improvements SO will offer to cold clump measurements in section~\ref{sec:so_fcast}. We conclude in section~\ref{sec:conclusion}.

\begin{figure}
\centering
\includegraphics[width=1\columnwidth]{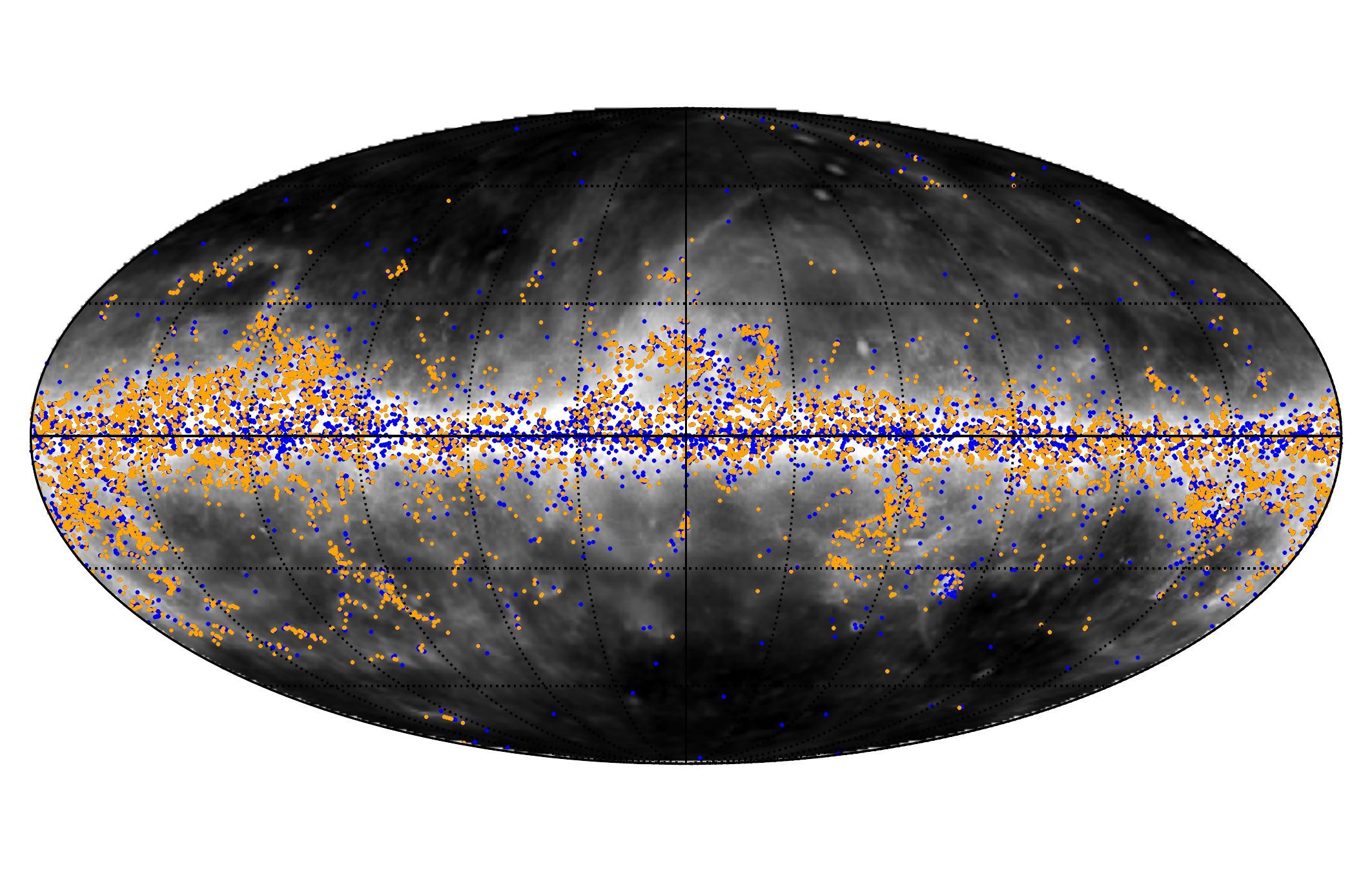}
\caption{All-sky distribution of the entire PGCC catalogue (blue) from \citet{planck2014-a37} and the subset used in this work (orange). The sources are overlaid on the 353$\,$GHz \textit{Planck} GNILC dust map, shown with normalized logarithmic scale between $5\times10^{-4}$ to 10$^{-2}$$\,$K$_{\mathrm{CMB}}$ \citep{planck2016-XLVIII}. The distribution of cold clumps is clearly concentrated at low and intermediate Galactic latitudes.\label{fig:allskyposns}}
\end{figure}


\section{Data Set}\label{sec:dataset}
In this section we introduce our sample selection from the PGCC Catalogue, the \textit{Planck} 353$\,$GHz full-sky maps required for stacking and the SO. 


\subsection{\textit{Planck} Galactic Cold Clump Catalogue}\label{sec:pgcc}
The \textit{Planck} catalogue of Galactic cold clumps \citep[PGCC; ][]{planck2014-a37} lists 13,242 Galactic objects with flux densities measured in three \textit{Planck} frequency bands (353, 545 and 857$\,$GHz) and the IRIS data from the Infrared Astronomical Satellite (IRAS) 3000$\,$GHz channel \citep{Miville-Deschenes2004, Neugebauer1984}. We can see these distributed across the entire sky as the blue dots in Fig.~\ref{fig:allskyposns}, overlaid on the \textit{Planck} GNILC thermal dust map \citep{planck2016-XLVIII}. The brightest sources are accompanied by spectral parameters such as the dust temperature, $T_{dust}$, and spectral index, $\beta$. 

The catalogue breaks the clumps into three categories based on the quality of their flux density values. We select our sample of cold clumps from the PGCC catalogue via the following criteria:
\begin{itemize}
    \item We use only the clumps labelled in the PGCC catalogue as \verb|FLUX_QUALITY=1|, corresponding to detection with accurate flux density estimates (S/N $>4$) in the \textit{Planck} 353, 545 and 857$\,$GHz bands as well as the IRIS 3$\,$THz channel. These clumps represent the highest quality clumps in the PGCC catalogue. 7012 of the 13,242 clumps in the PGCC catalogue are labelled as \verb|FLUX_QUALITY=1|.
    \item We use only the clumps flagged with \verb|FLUX_BLENDING=0| indicating that they are not directly overlapping one another. This cut ensures our map cutouts around each clump are centred on only one object to simplify the background correction for the stacking process. An additional 728 clumps fail the blending cut.
    \item We find two sources with peak $I^2$ and $P^2$ values over an order of magnitude larger than all other clumps. These correspond to clumps PGCC G15.02-0.67 and PGCC G351.45+0.67. Observing these objects in the \textit{Planck} maps we can see they are contaminated by nearby sources not flagged as other cold clumps and thus they weren't flagged by the above condition. We remove these two clumps.
\end{itemize}

Our final sample contains 6282 bright and reliable cold clumps. We show the locations of the full PGCC sample as blue dots in Fig.~\ref{fig:allskyposns}, and the subset used in this work as orange dots. 
For these 6282 cold clumps, \textit{Planck} reports luminosities for 2221 sources and reliable distance estimates for 209 clumps \citep[see][section 5 for their methods]{planck2014-a37}. We use these two subsets for the luminosity and distance tests in sections~\ref{sec:luminosity} and~\ref{sec:glatdist}.


\subsection{\textit{Planck} Maps}\label{sec:planck353}

\begin{figure}
\centering
\includegraphics[width=1\columnwidth]{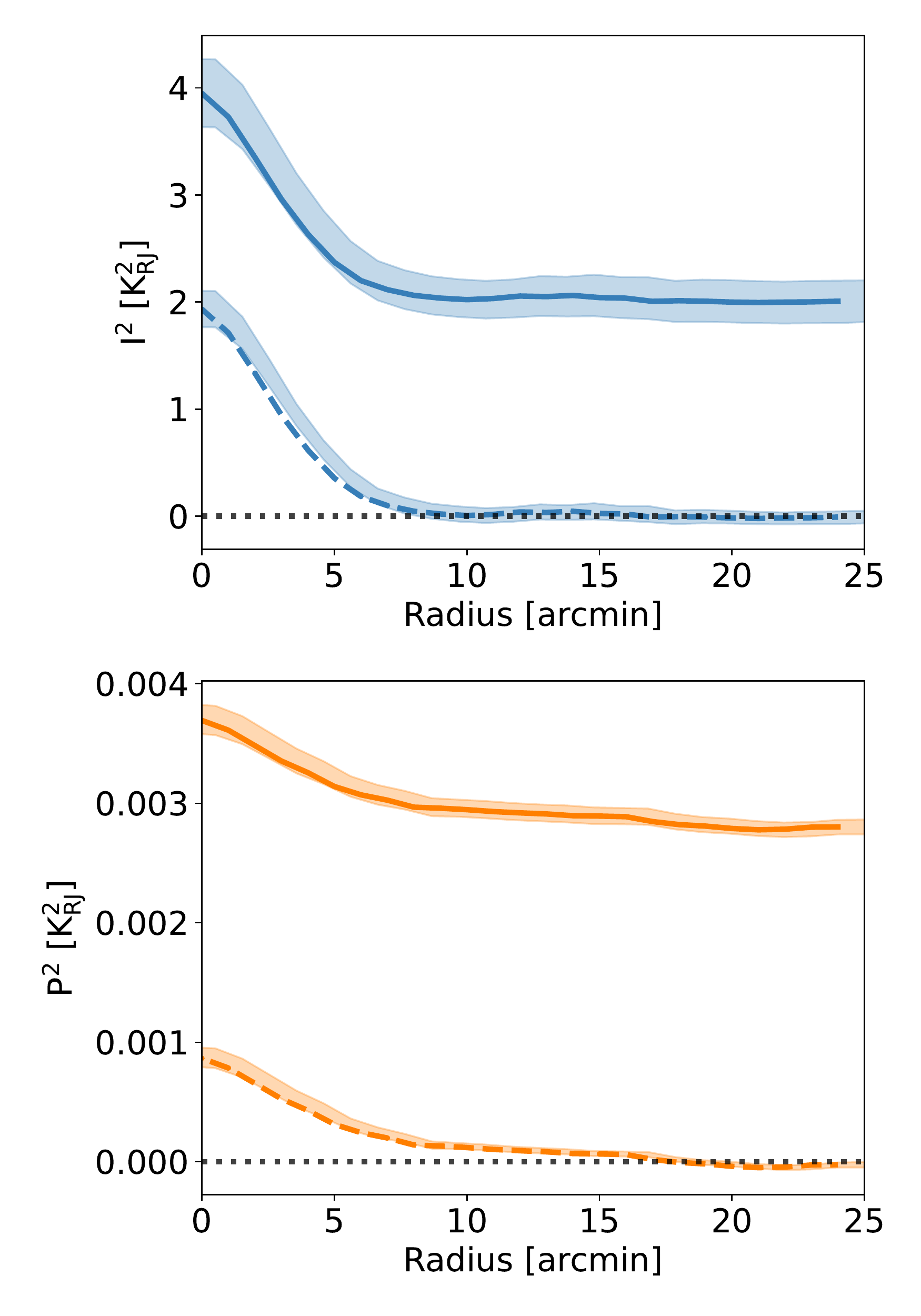}
\caption{Signal from 6282 PGCC sources stacked in squared intensity (top, blue) and polarization (bottom, orange) as a function of distance from the PGCC location. The solid lines denote the stacked result before bias correction. The dashed lines represent the processed signal with bias correction. The shaded regions are the 1$\,\sigma$ uncertainties from either stacked signal. Note by construction the bias correction is constructed to zero the signal from 15--24$\,$arcmin. We see a clear signal in both squared intensity and polarization (see section~\ref{sec:stacking}). \label{fig:IPStacks_1D}}
\end{figure}

We observe and later stack PGCC locations in the \textit{Planck} 2020 data release 353$\,$GHz maps, as these are the highest frequency maps with polarization information \citep{planck2020-LVII}. These full-sky, full-mission frequency products include improved high-resolution maps in three Stokes parameters; total intensity (\textit{I}), and linear polarization (\textit{Q} and \textit{U})\footnote{\url{http://pla.esac.esa.int/pla/aio/product-action?MAP.MAP_ID=HFI_SkyMap_353-ds3_2048_R4.00_full.fits}}.

The dominant signal in the 353$\,$GHz maps is thermal dust emission, although there are also very low levels of emission from other sources of emission such as the CMB \citep{planck2016-l03}. We chose the raw frequency maps over component separated maps \citep{planck2014-a12} to avoid biases or artifacts introduced by separation algorithms such as pixelization issues near bright sources or higher contamination of emission in different algorithms such as CIB in \verb|Commander| and \verb|GNILC| maps \citep{planck2016-l04}. 

These maps use the \verb|HEALPIX|\footnote{\url{http://healpix.sf.net/}} pixelisation \citep{Gorski2005, Zonca2019} with resolution $N_{side}=2048$ and have an effective beam of $\sim5$$\,$arcmin \citep{planck2016-l04}. 

A potential worry is the effect of any temperature-to-polarization leakage \citep{planck2020-LVII} that occurs on small angular scales. Temperature-to-polarization leakage would increase the polarization signal and hence the inferred polarization fraction.   
The PR4 maps have improved mitigation of systematic error, such as temperature and polarization leakage, compared to the earlier releases. As a test, we have compared our results with those derived from product release 3 (PR3) and find consistent outcomes. We assume the temperature-to-polarization leakage is negligible for the angular scales used in this work.

\defcitealias{Ade_2019}{SO19}

\subsection{The Simons Observatory}\label{subsec:so}
The SO is a new ground-based experiment optimized for CMB survey observations \citep[hereafter SO19]{Ade_2019}. It is currently under construction and will have three 0.5$\,$m telescopes and one 6$\,$m telescope at first light. The large-aperture telescope \citep[LAT;][]{Xu2021} will map 40$\,$per$\,$cent of the sky in six frequency bands (27, 39, 93, 145, 225 and 280$\,$GHz) and angular resolutions (8.4, 5.4, 2.0, 1.2, 0.9 and 0.8$\,$arcmin respectively) beginning in late 2023 \citep{Gudmundsson_2021}. The highest frequency band of the LAT (280$\,$GHz) has a projected sensitivity improvement of over five in intensity and over seven in polarization when compared to the \textit{Planck} 353$\,$GHz band \citep[\citetalias{Ade_2019}]{planck2016-l01}. The capabilities of SO will allow for sensitive measurements of Galactic emission to probe a wide set of science questions.

The SO science forecasts \citep[\citetalias{Ade_2019}]{Hensley_2022} predict an order of magnitude increase compared to \textit{Planck}, in the number of molecular clouds observed with at least 1$\,$pc resolution at $3\,\sigma$. SO will also be sensitive to the CO(2--1) transition line where, with low noise level and multi-frequency coverage, SO will constrain polarized CO emission at a level of polarization fractions $\sim$1$\,$per$\,$cent in the brightest molecular clouds. These surveys will provide a significantly larger sample of resolved Galactic cold clumps in polarized thermal dust and CO emission.  


\section{Methods}
\label{sec:Methods} 
We discuss now the method to estimate the polarization fraction of Galactic cold clumps via stacking, the procedure to account for noise and background contribution, and the error analysis.

From the \textit{Planck} PR4 \textit{I}, \textit{Q}, \textit{U} maps, we create two new maps, $I^2$ and $P^2 = Q^2 + U^2$. We use the squared maps so that the Galactic cold clump polarization is additive without needing to know the polarization direction. We then estimate the square of the polarization fraction, $\Pi^2$, defined as the ratio of the squared amplitude of linear polarization to total intensity: 

\begin{equation}\label{eqn:p2}
    \Pi^2\equiv\frac{P^2}{I^2} = \frac{Q^2+U^2}{I^2}.
\end{equation}

We will quote the mean squared and root-mean-squared (RMS) polarization fractions, defined as:

\begin{equation}\label{eqn:pfrac}
    \Pi^2_{MS} = \frac{\langle P^2\rangle}{\langle I^2\rangle} - N^2_{\mathrm{ring}},
\end{equation}

and 

\begin{equation}\label{eqn:RMSPF}
    \sqrt{\Pi^2_{MS}} = \sqrt{\frac{\langle P^2\rangle}{\langle I^2\rangle} - N^2_{\mathrm{ring}}}.
\end{equation}

where $N^2_{\mathrm{ring}}$ denotes a noise and background correction described in section~\ref{sec:noise_background}.
As noted earlier, we employ squared amplitude quantities in order to be able to stack without knowing the random polarization angles. This does introduce a challenge however, as the squared equations give positive-definite quantities: other sources such as noise have non-zero expectation values that will need to be subtracted to avoid bias. 
We discuss the treatment of bias terms in section~\ref{sec:noise_background}. 
We also note that we are estimating $\langle P^2\rangle/\langle I^2\rangle$. 
As motivated by \citet{Bonavera2017} and \citet{Trombetti2018} in similar stacking analyses of the polarization of extragalactic objects in \textit{Planck} maps, we choose this convention instead of $\langle P^2/I^2\rangle$ due to the low S/N on a single cold clump. Furthermore our choice allows for direct application to CMB foreground interpretations where the power spectrum of sources in polarization follow a $P^2$, power-weighting dependence analogous to our measured quantities.

\subsection{Stacking}\label{sec:stacking}
We perform a stacking procedure to obtain a measurement of the mean-squared and RMS polarization fraction, similar to other stacking analyses \citep[e.g.][]{Montier2005,Dole2006,Bethermin2012,Bonavera2017, Gupta2019}. This statistical method consists of adding up many regions of the sky centred at selected positions to enhance a signal we wish to observe. 
In particular, this is useful when sources within the sample are too faint to be detected individually at the desired frequency. We are able to measure their mean-squared polarized flux density despite individual sources not being directly detectable in polarization. 

We extract a square $50\times50$$\,$arcmin patch centred at the location of each cold clump with a pixel size of 1$\,$arcmin from the $P^2$ and $I^2$ maps. This yields 6282 square maps (one per cold clump in the sample), which are averaged to create a mean polarization $\langle P^2\rangle$ and intensity $\langle I^2\rangle$ map. Given some range in physical clump size, we chose the patch size such that we could project the smallest clumps to appropriately stack their peak, and the largest would fit with sufficient distance from the edges to perform background measurements. 
Given our sample criteria removes any sources with separations $\le15$$\,$arcmin it is very uncommon for multiple PGCC's to be found within the same patch, however there may still be part or all of another clump within the background of the target source. These may serve as an additional background contaminant to suppress the total signal after background subtraction (section~\ref{sec:noise_background}) however the contribution would be minimal given these themselves would be suppressed into the background in the stacking.

In Fig.~\ref{fig:IPStacks_1D} we plot the mean-squared signal from the stacked PGCCs in $I^2$ (top) and $P^2$ (bottom) as a function of radius from the peak signal at the centre of the stack. 
We observe a clear signal in both total intensity and polarization. The solid lines demonstrate the positive noise bias due to using squared quantities, as we can see the level does not asymptote to zero at large distances. A correction from this background and noise (discussed below in section~\ref{sec:noise_background}) has been applied to the dashed lines, where we can see the background signal has been reduced to be mean zero by $\sim$15$\,$arcmin from the signal centre in both intensity and polarization. 

We take the mean of the background-subtraction $I^2$ ($P^2$) map within 3$\,$arcmin of the centre of the mean-squared intensity (polarization) signal. We then take the ratio of the $\langle P^2\rangle/\langle I^2\rangle$ to report the measured mean-squared polarization fraction.


\subsection{Noise \& Background Correction}\label{sec:noise_background}

The measured values of $\langle P^2\rangle$ and $\langle I^2\rangle$ at cold clump locations will be biased high by signals and noise in the \textit{Planck} maps. These terms include noise, diffuse emission from our Galaxy, and the CMB signal (sub-dominant at 353$\,$GHz). 

Although stacking enhances the signal-to-noise, the positive-definite nature of $I^2$ and $P^2$ means that the contribution from noise and other signals will bias high estimates of the thermal dust polarization and total intensity. This bias is especially significant for $P^2$ where S/N is significantly lower. We need to measure and subtract this background contribution external to our stacked signal.

To estimate the contribution of background and remaining noise, we define a combined noisy and background contaminated estimate $X'$ of each Stokes parameter $X\in[I,Q,U]$,

\begin{equation}\label{eqn:noisyestimate}
    X'=X+N_X,
\end{equation}

where $N_X$ is a contribution from noise and background contamination. From equation~\ref{eqn:p2}, we need an unbiased estimate of $X^2$ which we can write as:

\begin{equation}\label{eqn:backgroundsubtraction}
    X^2 = X'^2 - N_X^2.
\end{equation}

To solve for $N_X^2$ we estimate the background signal of our total summed patches of $I^2$ and $P^2$ by measuring the mean signal in an annulus with inner radius 18$\,$arcmin and outer radius of 24$\,$arcmin ($X_{\textrm{ring}}$),

\begin{equation}
    N_X^2 = \langle X^2_{\textrm{ring}}\rangle,
\end{equation}

where these radii were chosen such that the ring contains nearly all the background of the patch whilst fitting inside the 50$\times$50$\,$arcmin square cutout and avoiding the cold clump signal.

As in equation~\ref{eqn:backgroundsubtraction}, we subtract this average stacked background signal in $I^2$ and $P^2$ from the final respective stacks (Fig.~\ref{fig:IPStacks_1D} [dashed lines]). We acknowledge there will likely remain a bias in the form of a cross term of the background and signal from the square of equation~\ref{eqn:noisyestimate}. This, however, should be small and we are unable to remove it due to using positive-definite values. These terms are fluctuating about zero, requiring us to perform the square to positive-definite, performing the annulus background subtraction before would not guarantee the bias is removed.

We compare a second background and noise correction method as a consistency check in which we apply a high-pass filter and subtract an average simulated noise signal from the \textit{Planck} IQU maps. This method more directly targets large scale background structure and small scale noise contributions, although it doesn't account for other Galactic and extra-Galactic signals at smaller scales. 

The filter is applied to the \textit{Planck} maps at an angular scale of $\ell=400$, smoothing out all structure larger than the largest cold clumps. We also apply this to a set of 100 independent noise realizations from the PR4 simulations. These simulations are  maps of noise and systematic residuals only (CMB and foregrounds have been subtracted), and are available on the \textit{Planck} Legacy Archive\footnote{Example noise realisation: \url{http://pla.esac.esa.int/pla/aio/product-action?SIMULATED_MAP.FILE_ID=ffp10_noise_100_full_map_mc_00000.fits}} (PLA). We take the mean of the filtered noise and subtract from the filtered $I^2$ and $P^2$ maps. 

With the filtered version, we find higher values of $I^2$ and $P^2$, along with a non-zero background level far away from the source. These effects are presumably due to the contributions of other signals in the maps at small angular scales. As the bias is fractionally larger in $P$ than $I$, this leads to a larger apparent S/N (8.6$\,$$\sigma$) on the mean-squared polarization fraction. However, we attribute this to the residual background bias and only report results for the annulus subtraction technique in the rest of this work.

The dashed lines of Fig.~\ref{fig:IPStacks_1D} demonstrate the stacked signal corrected using the annulus background subtraction method. Comparing the uncorrected results (solid lines), we see a significant increase in S/N where the backgrounds in $P^2$ and $I^2$ are reduced to zero $\sim$15$\,$arcmin from the signal centre. Polarization remains comparatively noisier however the background subtraction is a significant improvement when measuring the peak signal.


\subsection{Error Estimation}\label{sec:error}
To estimate the uncertainties on $\Pi^2_{MS}$, we use a bootstrapping technique of resampling with replacement \citep{Gupta2019}. 
We randomly sample from the full set of $N_c=6282$ objects\footnote{We use the same procedure when quoting errors on the subsets of the full sample.}, measure $\langle P^2\rangle$ and $\langle I^2\rangle$ for the random sample of cold clumps, and repeat this 2500 times to obtain

\begin{equation}
\mu_{\Pi^2} = \frac{\mu_{P^2}}{\mu_{I^2}},
\end{equation}

where $\mu_P = \sum_{i}\langle P^2\rangle_i/N_c$ and $\mu_I = \sum_{i}\langle I^2\rangle_i/N_c$.
The standard deviation of the resulting mean $\Pi^2_{MS}$ values is taken to be the uncertainty on the $\Pi^2_{MS}$ measurement for the sample of clumps:

\begin{equation}
    \sigma_{\Pi^2} = \sqrt{\frac{\sum_i{(\Pi^2_{MS,i} - \mu_{\Pi^2})^2}}{N_c}}.
\end{equation}

We can then obtain the uncertainty on the RMS polarization fraction as:

\begin{equation}
    \sigma_{\sqrt{\Pi^2}} = \frac{\sqrt{\Pi^2_{MS}}}{2}\frac{\sigma_{\Pi^2}}{\Pi^2_{MS}}.
\end{equation}

These uncertainties includes both population and measurement error.


\section{Results}\label{sec:results}

We detect the stacked PGCC signal at high S/N in both intensity and polarization, as shown in Fig.~\ref{fig:IPStacks_1D}. The squared intensity peaks at $\langle I^2\rangle \approx 1.9$$\,$K$^2_{\mathrm{RJ}}$, while the squared polarization signal peaks at $ \langle P^2\rangle \approx 9e-4\,$K$^2_{\mathrm{RJ}}$. 
We measure the mean-squared polarization fraction within a $3\,$arcmin radius of the centre to be $\Pi^2_{MS}=[4.79\pm0.44]\times10^{-4}$ with an 11$\,\sigma$ detection, corresponding to an RMS polarization fraction $\sqrt{\Pi^2_{MS}}=[2.19\pm0.10]\,$per$\,$cent. The clear detection of polarization validates the potential of mm-wavelength polarization observations to study magnetic fields within cold clumps.

Considering the RMS polarization fraction as a proxy of the average polarization fraction over the population\footnote{Whilst this is a fair approximation it should be manipulated carefully.}, the observed polarization fraction in these clumps is significantly lower than \citet{planck2015-XXXVIII} 
found in an analysis of filamentary structures at high Galactic latitudes in the same maps. 
The filament analysis was based on 259 identified filaments with $|b|\ge2^\circ$. 
Using a stacking analysis instead over $I$, $Q$ and $U$, with $Q$ and $U$ aligned along the filaments long axes, they find the mean polarization fraction to be $11$$\,$per$\,$cent. A possible explanation of this discrepancy could be their avoidance of using $P^2$ and the introduction of squared bias amplitudes, however, without a common alignment strategy such as aligning the cutouts along the long axis of filaments, this method is not possible. Another explanation is that the lower polarization fraction in the cold clumps is due to a more complex magnetic field structure in the clumps, leading to increased depolarization due to the different magnetic field orientations along the line of sight passing through the clump.
Interestingly, the observed cold clump polarization fraction in this work also increases at the same latitudes, $|b| > 30^\circ$, as the filament sample. 
The polarization of cold clumps at different latitudes is discussed in section~\ref{sec:glatdist}.

Given the 11$\,\sigma$ measurement of the mean-squared polarization fraction of cold clumps, we consider splitting the PGCC sample to study possible trends of polarization with key parameters of the cold clump distribution, structure and composition. We will look for relationships between polarization fraction and total flux density, luminosity and Galactic latitude. 
As the specific clumps sample used varies between some splits, we also compile the mean squared polarization fraction and RMS polarization fraction for each subsample in Table~\ref{table:allresults}.
The measured mean-squared and RMS polarization fraction are similar across all subsamples used in the following subsections. 

\begin{table}
\begin{tabular}{p{12.5em} c c}
    \hline
    Sample used                                 & $\Pi^2_{MS}$ \small{$\left(\times10^{-4}\right)$} & $\sqrt{\Pi^2_{MS}}$ \small{$\left(\%\right)$} \\
    \hline
    Full sample                                 & 4.79~$\pm$~0.44                                            & 2.19~$\pm$~0.10 \\
    Total flux density bin average              & 4.87~$\pm$~0.35                                            & 2.22~$\pm$~0.08 \\
    Luminosity bin average                      & 4.69~$\pm$~0.54                                            & 2.19~$\pm$~0.13 \\
    Galactic latitude bin average               & 5.19~$\pm$~0.39                                            & 2.36~$\pm$~0.09 \\
    Distance bin average [High b]               & 4.14~$\pm$~0.57                                            & 2.04~$\pm$~0.14 \\
    Distance bin average [Low b]                & 4.11~$\pm$~0.54                                            & 2.07~$\pm$~0.13 \\
    \hline
\end{tabular}
\caption{A summary of the bin-average mean-squared and RMS polarization fraction results from each test to 
show consistency across the different subsamples of cold clumps. 
All results have been calculated following the stacking procedure in section~\ref{sec:stacking} with annulus background subtraction and error calculated as per section~\ref{sec:error}.}
\label{table:allresults}
\end{table}


\subsection{Flux Density}\label{sec:flux_density}

More distant cold clumps will naturally tend to have lower flux densities, $S$.
As  distance effects are expected to dominate the observed flux distribution, we do not expect to observe a correlation between the measured polarization fraction and flux density \citep[e.g.][]{Puglisi_2018}. 
We test this expectation by splitting the 6282 cold clumps into six flux bins based on their total flux density as reported by \textit{Planck}.
We set the bin edges to achieve approximately equal signal-to-noise in intensity. 
The bins are defined as 0--1.38, 1.38--2.35, 2.35--5, 5--7.4, 7.4--12.2 and 12.2--320$\,$Jy, and contain 539, 928, 1990, 1014, 909 and 902 clumps respectively.

\begin{figure*}
\centering
\includegraphics[width=0.65\textwidth]{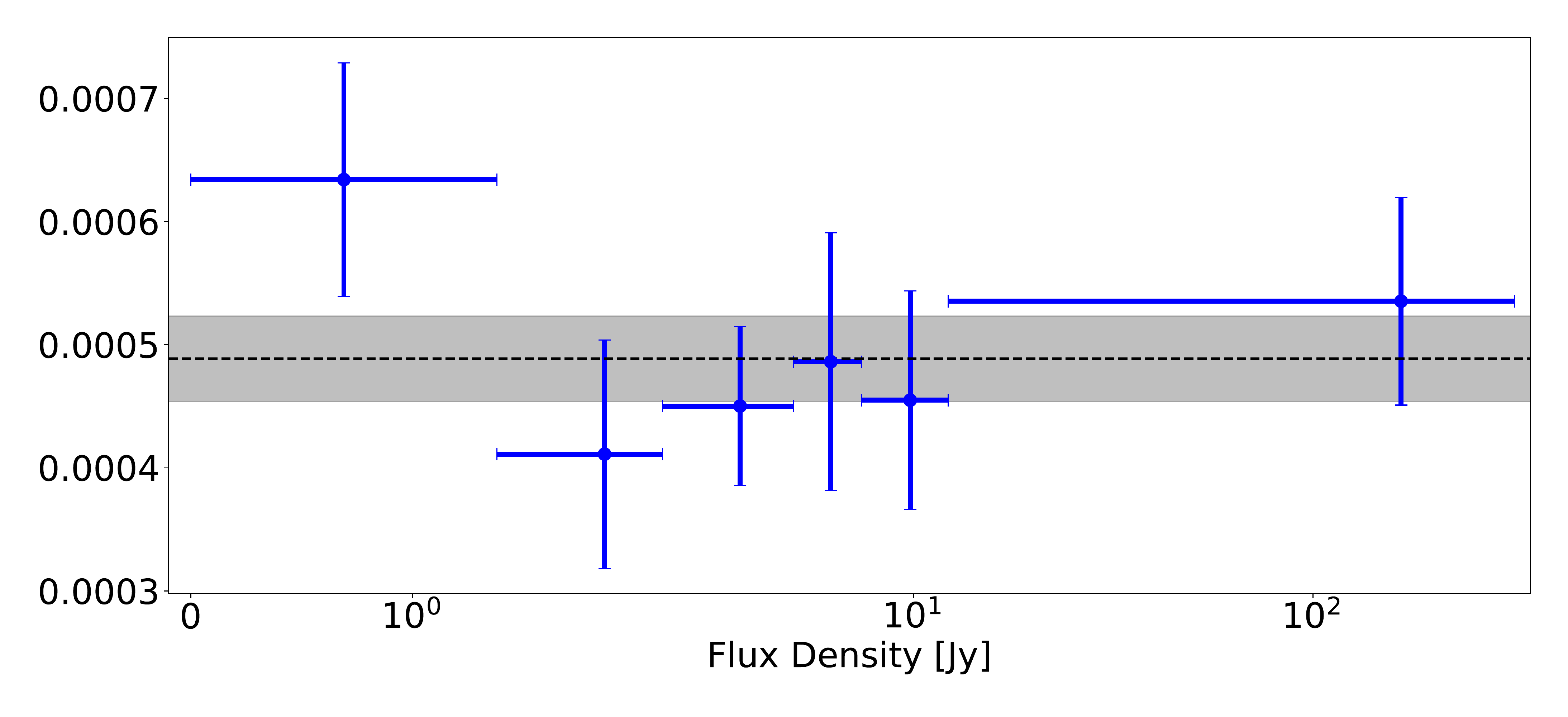}
\caption{Mean-squared polarization fraction measured in six bins of increasing flux density ranges (Jy). The vertical error bars are calculated as per section~\ref{sec:error}. The horizontal dashed line represents the error-weighted bin-average mean-squared polarization and the gray region being the corresponding error ($\Pi^2_{MS}=[4.87\pm0.35]\times10^{-4}$). We see no indication the polarization of Galactic cold clumps is dependent on their flux density.\label{fig:fluxbins}}
\end{figure*}

Figure~\ref{fig:fluxbins} shows the resulting mean-squared polarization fraction in each flux bin. 
Each bin's polarization fraction is calculated following the same stacking method as in section~\ref{sec:Methods}, and the error bars are calculated as in section~\ref{sec:error}.
The horizontal dotted line represents the error-weighted average mean-squared polarization fraction from all bins to check for consistency with the full sample result, where we measure $[4.87\pm0.35]\times10^{-4}$ and the error in this is given by the shaded gray region (the corresponding RMS result is listed in Table~\ref{table:allresults}). 
As expected, there is no apparent trend with flux density. 
We confirm this by performing a linear fit between flux density and mean-squared polarization fraction. 
The recovered slope is [$0.03\pm0.05$]$\times10^{-5}$, consistent with zero at $0.6\,\sigma$. 
There is no evidence that the mean-squared polarization fraction depends on the flux density of a cold clump.

\subsection{Luminosity}\label{sec:luminosity}

\begin{figure}
\centering
\includegraphics[width=0.4\textwidth]{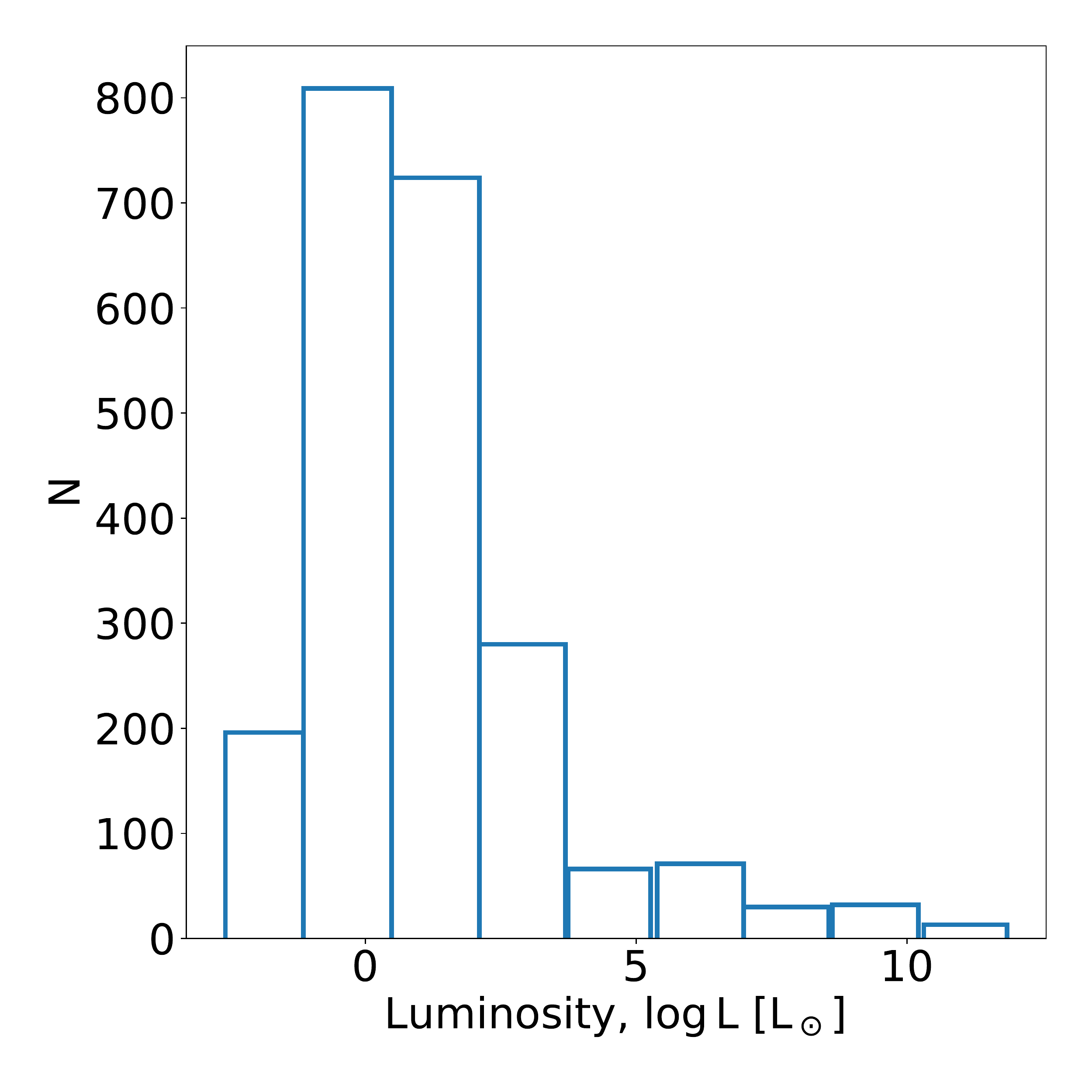}
\caption{Histogram of 2221 PGCCs split into nine logarithmic luminosity bins. The bins contain 196, 809, 724, 280, 66, 71, 30, 32 and 13 cold clumps respectively. The distribution of Galactic cold clumps extends over orders of magnitudes of emitted power, but are heavily weighted towards the L$\in[10^0,10^2]$L$_\odot$ range. \label{fig:luminosity_hist}}
\end{figure}

As with flux density, we do not expect to find a correlation between the luminosity, L, and polarization fraction of cold clumps. 
Luminosity estimates are not available for the full PGCC sample; \textit{Planck} reports luminosities for a sub-sample of 2221 cold clumps \citep{planck2014-a37}.
As with flux density, we split these 2221 cold clumps into nine logarithmic luminosity bins. 
The bin edges are $6.3\times10^{-2}$--$0.32$, $0.32$--$1.6$, $1.6$--$9.3$, $9.3$--$42$, $42$--$2.1\times10^{2}$, $2.1\times10^{2}$--$1.1\times10^{3}$, $1.1\times10^{3}$--$5.5\times10^{3}$, $5.5\times10^{3}$--$2.8\times10^{4}$ and $2.8\times10^{4}$--$1.4\times10^{5}\,$L$_\odot$. The number of Galactic cold clumps in each bin is shown in Fig.~\ref{fig:luminosity_hist}.

\begin{figure*}
\centering
\includegraphics[width=0.65\textwidth]{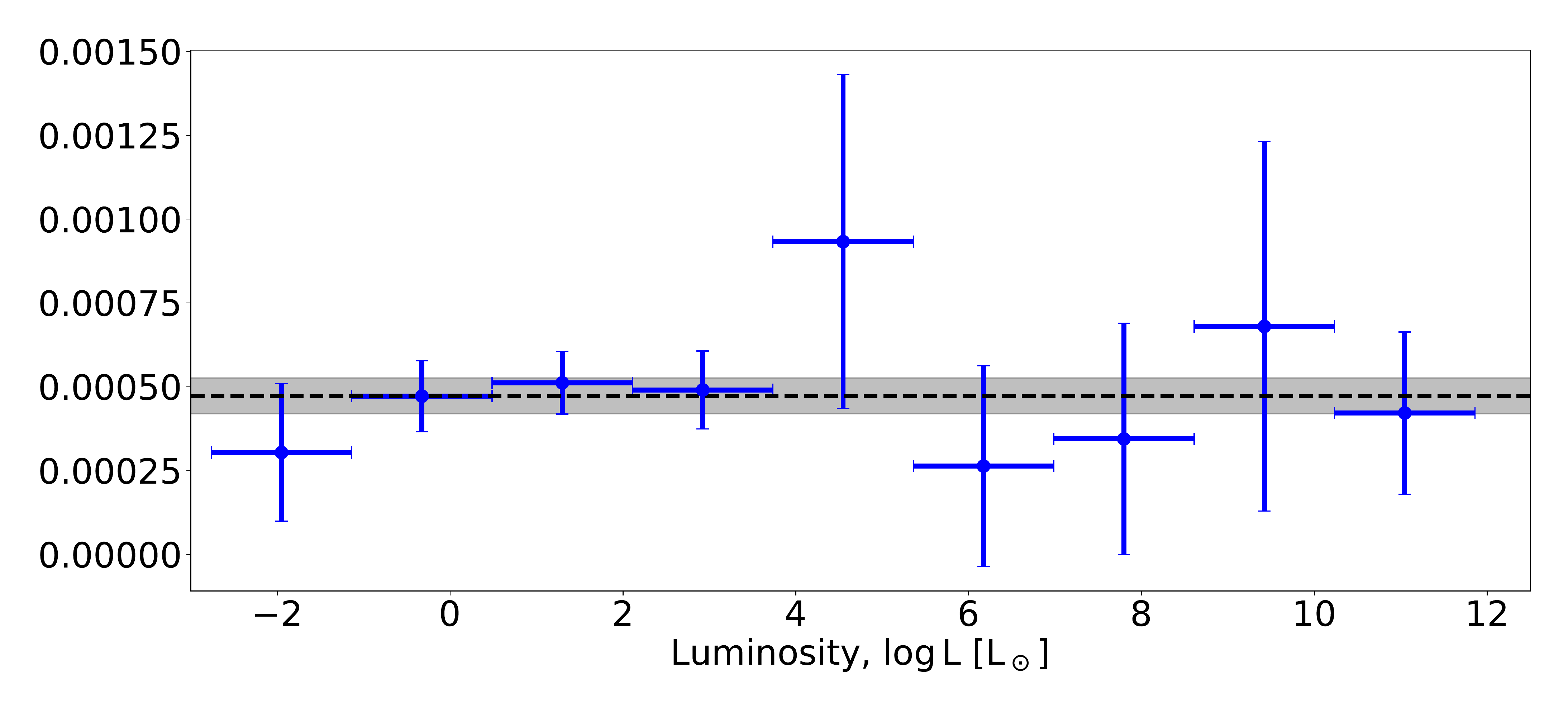}
\caption{Mean-squared polarization fraction measured in nine logarithmic bins of increasing luminosity. The horizontal black dashed line represents the error-weighted bin-average mean-squared polarization, $\Pi^2_{MS}=[4.69\pm0.54]\times10^{-4}$, where the gray region is the corresponding error. There is no sign that the polarization fraction depends on the luminosity of cold clumps. Vertical error bars are calculated as per section~\ref{sec:error} and the horizontal error bars are the bin ranges.\label{fig:luminosity_scatter}}
\end{figure*}

In Fig.~\ref{fig:luminosity_scatter}, we show the measured mean-squared polarization fraction estimated for each stacked luminosity bin. 
The horizontal dashed line indicates the weighted average mean-squared polarization fraction across the bins, $[4.69\pm0.54]\times10^{-4}$, with the gray shaded region showing the $1\,\sigma$ interval.
The weighted average estimates for the mean-squared polarization and RMS polarization fraction are also listed in Table~\ref{table:allresults}. 
As with flux density, there is no sign of a trend. 
We fit for a linear relationship between the luminosity and mean-squared polarization fraction, finding $[0.006\pm10.9]\times10^{-6}\mathrm{ L }+[4.73\pm0.39]\times10^{-4}$.
The slope is consistent with zero at $\ll1\,\sigma$.
We conclude that there is no evidence that the mean-squared polarization fraction depends on the luminosity of a cold clump.


\subsection{Galactic Distribution}\label{sec:glatdist}

\begin{figure}
\centering
\includegraphics[width=0.4\textwidth]{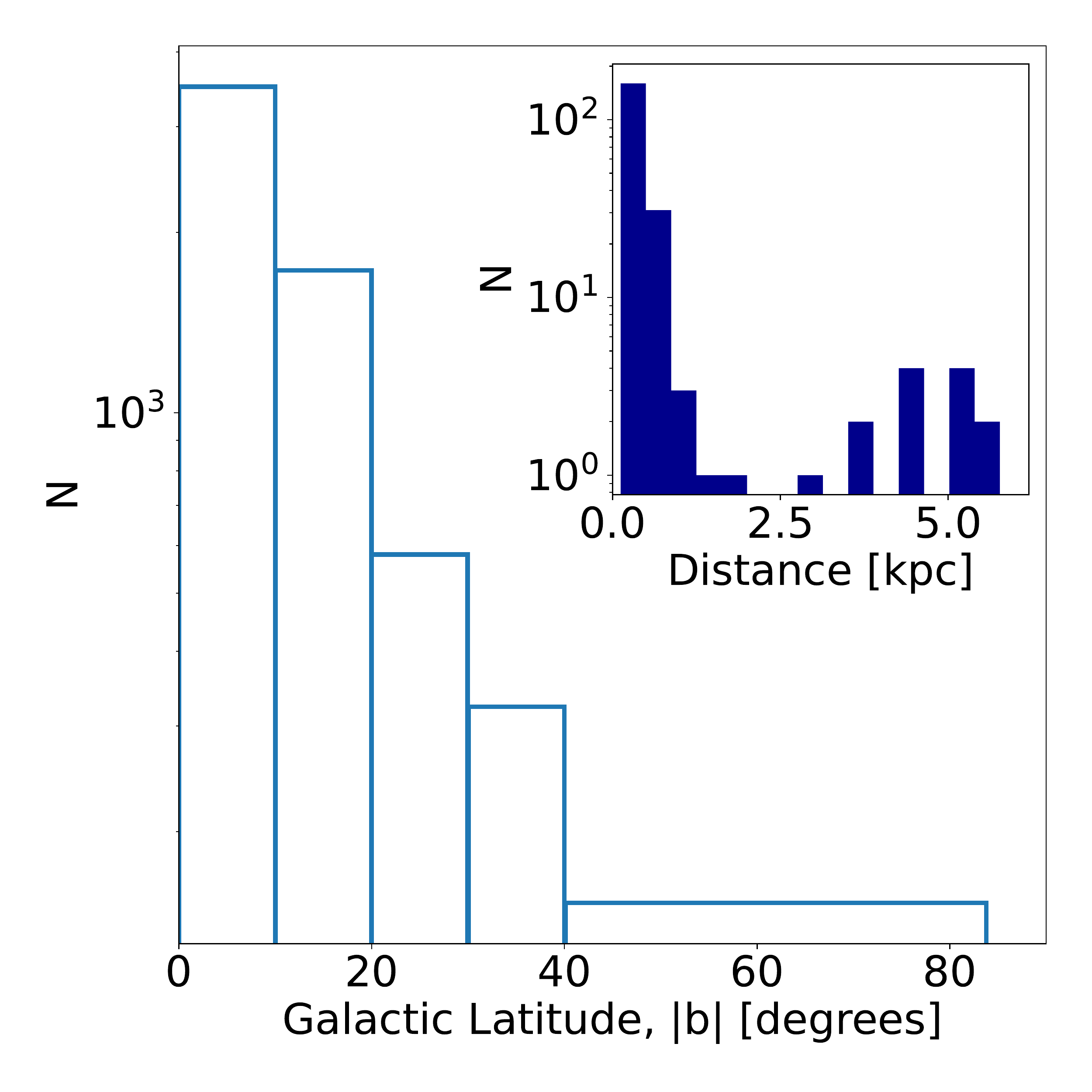}
\caption{Histogram of PGCCs split by their Galactic latitude into four $10^\circ$ bins from 0--40$^\circ$ and one $50^\circ$ bin from 40--90$^\circ$. The bins contain 3499, 1728, 580, 323 and 152 cold clumps respectively. The clumps are concentrated about the Galactic midplane but do extend out to high galactic latitudes. [Inset] Histogram of PGCCs with reliable distance measurements ranging from $\sim70\,$pc to $\sim6\,$kpc.  \label{fig:glathist}}
\end{figure}

We next test if the mean-squared polarization fraction of PGCCs varies with their Galactic latitude or their physical distance to the Sun. 
These tests could offer an insight into the importance of environmental effects on the polarization properties and magnetic field structure of Galactic cold clumps.

As in other tests, we first split the sample of cold clumps into bins based on the quantity of interest, in this case the magnitude of a clump's Galactic latitude. 
We use four equal-width bins from $0^\circ$ to $40^\circ$, and a fifth bin for Galactic latitudes from $40^\circ$ to $90^\circ$. 
The number of cold clumps in each bin is plotted in the main panel of Fig.~\ref{fig:glathist}. 
The majority of the clumps in the sample are concentrated at low Galactic latitudes, and thus we  expect (and find) the stacked measurement to be more uncertain at high latitudes.

\begin{figure*}
\centering
\includegraphics[width=0.65\textwidth]{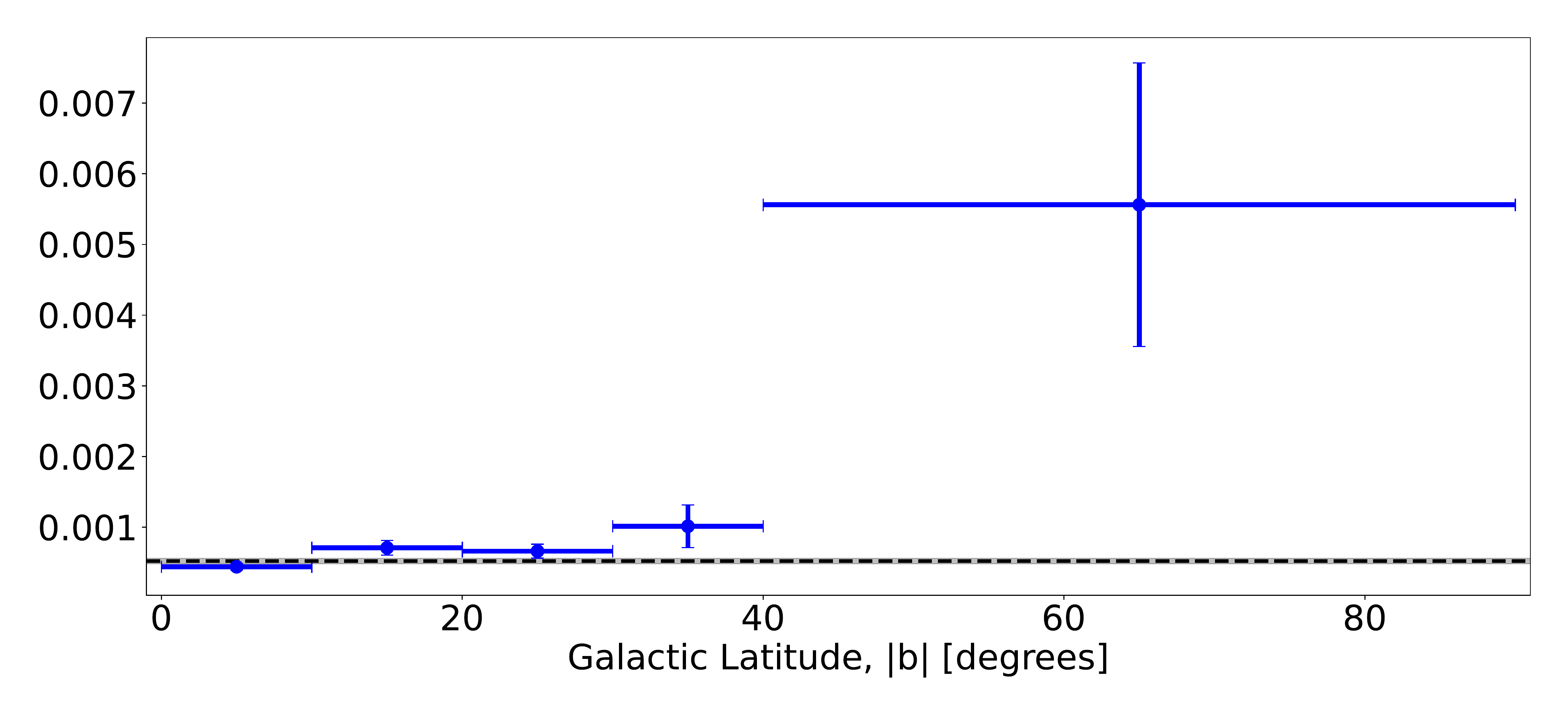}
\caption{Mean-squared polarization fraction calculated with our full sample of cold clumps binned by Galactic latitude. The points represent the noise and background corrected results in each bin. The vertical error bars are calculated as per section~\ref{sec:error} and the horizontal bars represent the bin regions. The horizontal dashed line represents the error-weighted bin-average mean-squared polarization fraction, $[5.19\pm0.39]\times10^{-4}$, where the error is represented by the gray band. The increase in polarization fraction with Galactic latitude follows that of the total Galactic dust behaviour as discussed in section~\ref{sec:glatdist}. This may be due to residual background contributions or distance effects (Figs.~\ref{fig:lowbdist}\,\&\,\ref{fig:highbdist}) where we are statistically limited to draw strong conclusions. \label{fig:lowvshigh_glat}}
\end{figure*}

We see a clear increase in the mean-squared polarization fraction at high Galactic latitudes, as shown in Fig.~\ref{fig:lowvshigh_glat}.
This increase is especially significant in the highest bin for $|b| > 40^\circ$, which at $[5.56\pm2.00]\times10^{-3}$ is $\sim25\,\sigma$ higher than the full sample mean-squared polarization fraction of $[4.79\pm0.0.44]\times10^{-4}$. 
The apparent increase might be real and indicate an environmental effect on the magnetic field structure of the cold clumps. 
Alternatively, it might indicate a bias in the background subtraction. 
The line-of-sight to low Galactic latitude clumps will tend to pass through a higher column density of dust. 
These higher column densities will depolarize the total signal before background subtraction \citep{Myers1991}. 
The increase in polarization fraction with decreasing total gas column density was also observed by \citet{planck2016-l11B}.
While we would expect the background subtraction of section \ref{sec:noise_background} to remove this effect, imperfect  background subtraction might thus explain the observed trend. 
A larger sample of cold clumps at high Galactic latitudes would help distinguish between these alternatives.

\begin{figure*}
\centering
\includegraphics[width=0.65\textwidth]{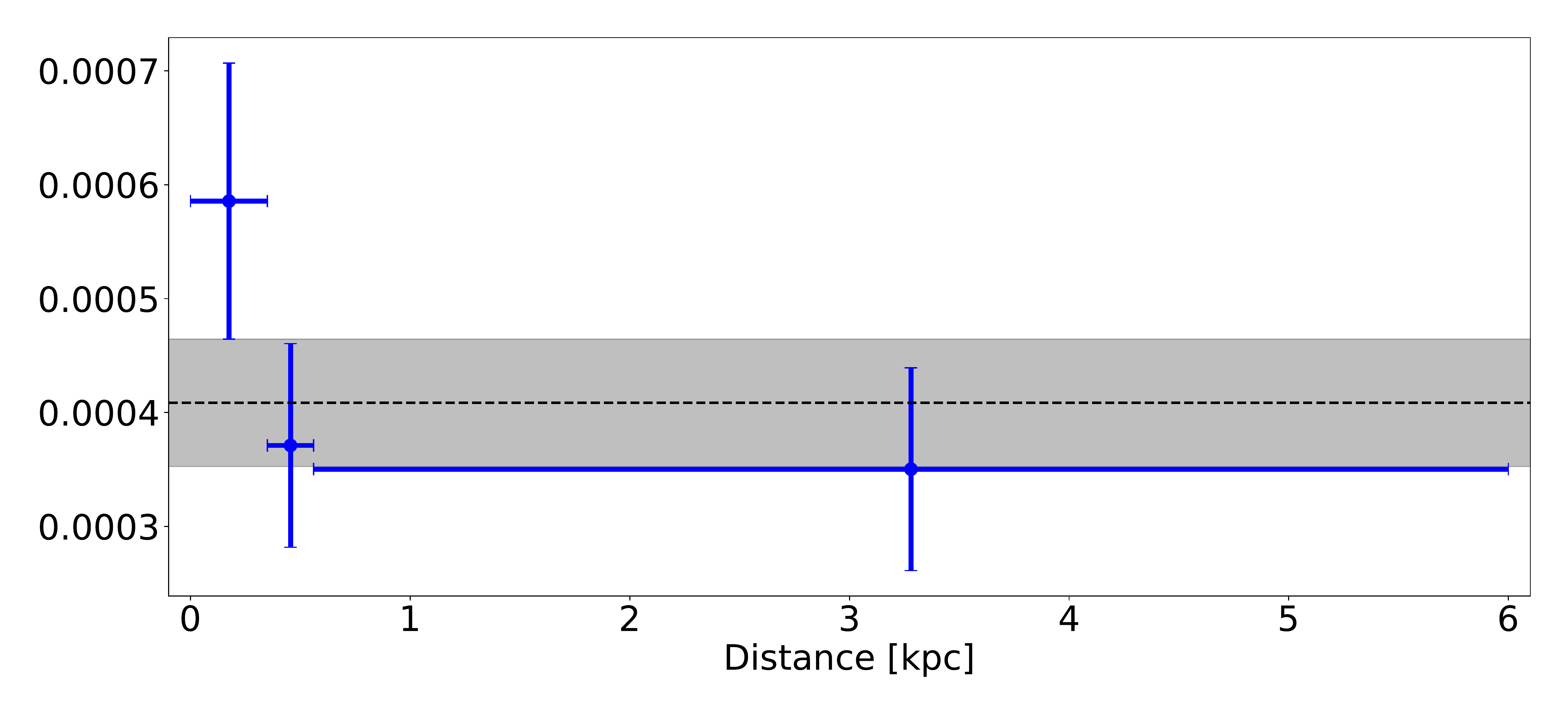}
\caption{Mean-squared polarization fraction for a sample of cold clumps with reliable distances binned according to their physical distance at low Galactic latitude ($|b|<10$). The vertical error bars are calculated as per section~\ref{sec:error} and the horizontal bars represent the bin regions. The horizontal dashed line represents the error-weighted bin-average mean-squared polarization fraction, $[4.11\pm0.54]\times10^{-4}$, where the error is represented by the gray band. We find no dependence of polarization fraction on distance at low Galactic latitudes; the slope of the linear fit is consistent with zero at $\lesssim1$\,$\sigma$. \label{fig:lowbdist}}
\end{figure*}

\begin{figure*}
\centering
\includegraphics[width=0.65\textwidth]{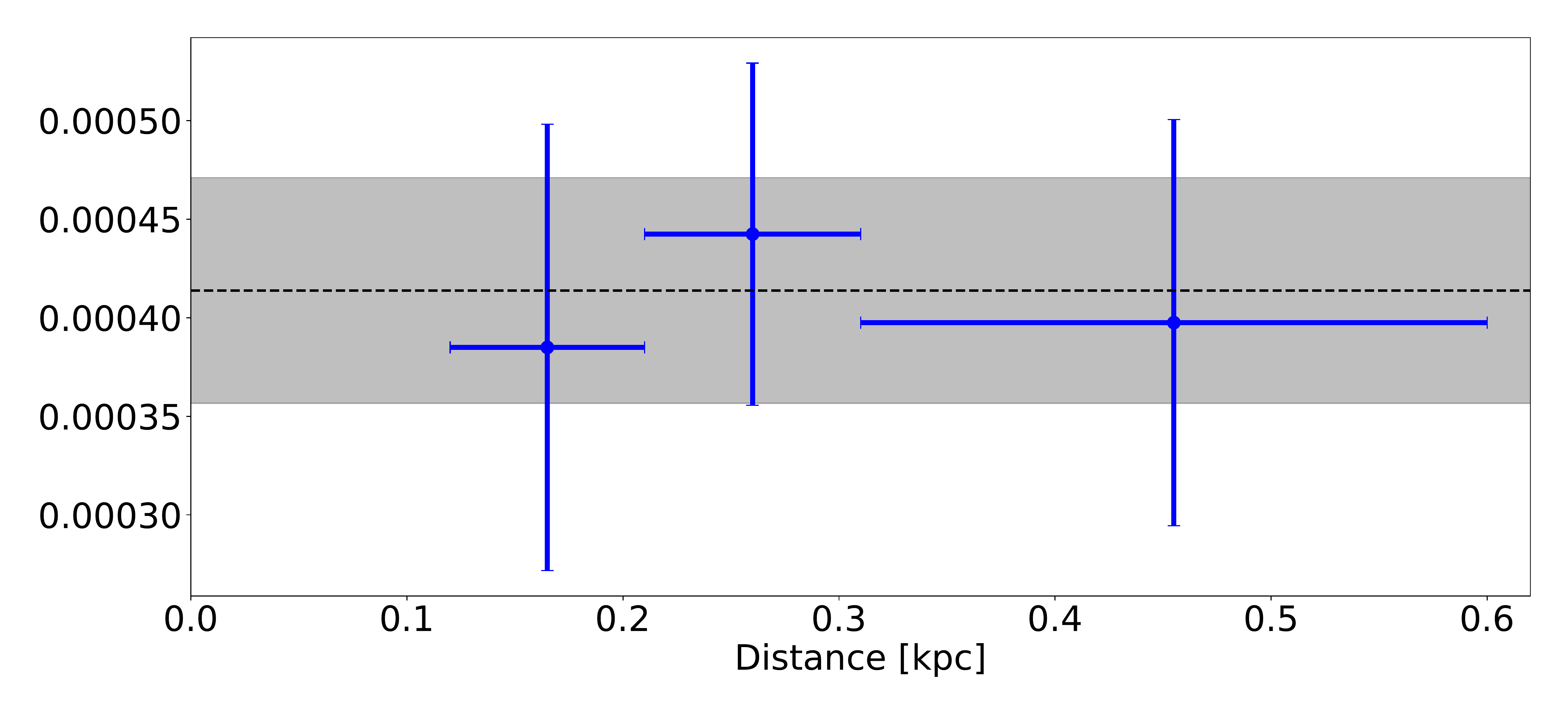}
\caption{Mean-squared polarization fraction for a sample of cold clumps with reliable distances binned according to their physical distance at high Galactic latitude ($|b|>10$). The vertical error bars are calculated as per section~\ref{sec:error} while the horizontal bars represent the bin regions. The horizontal dashed line represents the error-weighted bin-average mean-squared polarization fraction, $[4.14\pm0.57]\times10^{-4}$, where the error is represented by the gray band. We find no dependence of polarization fraction on distance at high Galactic latitudes; the linear fit slope is consistent with zero at $\lesssim1$\,$\sigma$. \label{fig:highbdist}}
\end{figure*}

Finally, we split the cold clump sample based on their physical distance to the Sun. 
\citet{planck2014-a37} estimate reliable distances for a subset of the full sample, as noted in section~\ref{sec:pgcc}.
The distance distribution for this sample is shown in the inset to Fig.~\ref{fig:glathist}. 
Nearly all clumps are located below 1$\,$kpc, though a small number extend out to $\sim6\,$kpc.
Note that the subsample  with reliable distances tends to be closer and at higher flux than the full sample. 
We split the reliable distance sample first into high and low Galactic latitudes, splitting at $|b| = 10^\circ$. 
We then further split each set into three distance bins. 
The mean-squared polarization fraction for the high Galactic latitude sample across the distance bins is shown in Fig.~\ref{fig:highbdist}, and the low Galactic latitude sample  in Fig.~\ref{fig:lowbdist}.
The average mean-squared polarization fraction is consistent between the high and low Galactic latitude sample, at $4.14\pm0.57$ versus $4.11\pm0.54$.
Neither subsample show strong evidence for a trend in the polarization fraction with distance. 
However, the high latitude subsample is limited to very nearby clumps as the maximum distance for that sample is $\sim0.6\,$kpc. 
The improved sensitivity and angular resolution of SO will allow cold clumps to be detected at larger distances and allow a much wider range of distances to be tested.

\subsection{Galactic Cold Clump Masks}


We are releasing full-sky masks of Galactic cold clumps, publicly available in Stokes \textit{I}, \textit{Q} and \textit{U}. CMB experiments aim towards many key science goals that will require dusty foreground objects such as PGCCs to be masked as to avoid contamination in power spectrum analyses. For example, SO aims to make a significant contribution to the detection and constraint of primordial \textit{B}-mode polarization signals. These signals are imprinted at angular scales between 10$\,$arcmin and a few tens of degrees during the epoch of inflation \citep{Seljak1996, Kamionkowski1996} and hence present at scales at which Galactic emission is a formidable foreground component. This Galactic emission ranges from synchrotron radiation to thermal dust emission \citep{Page2007, Bennett2013, planck2014-a12, planck2016-l01} and considering the scales at which these \textit{B}-modes are imprinted, PGCCs in particular may pose as an important component of the Galactic thermal dust emission requiring careful removal. 

As discussed in section~\ref{sec:intro} the PGCC catalogue lists the Galactic coordinates of each core along with some basic size information. Particularly useful to us is the listed full width at half maximum (FWHM) major and minor axes calculated for each clump. With these, along with an observed position angle, we are able to construct a 2D elliptical Gaussian profile for each PGCC, centred at the sources Galactic coordinates. The angular pixel and vector locations to position the profiles are obtained using the \verb|healpy| python package \citep{Zonca2019} on the HEALPix coordinates at the PGCC positions. 

The elliptical Gaussian masks also feature a user defined apodization threshold. This allows some flexibility to adjust how harsh the mask acts, smaller apodization will result in a slightly larger mask but may reduce potential residual signal from the clump. Setting this to 0 would result in a circular and binary mask if required. Given a circular, binary mask, an extent setting is applied defining the radius of the circular mask as a multiple of the clumps FWHM major axis. By default this is set at $1.5\times\textrm{FWHM}_{\textrm{maj}}$.

These masks (based on the full PGCC catalogue) are publicly available. We have also made the Python functions used to create these masks available so users can generate their own masks with different requirements such as resolution ($N_{side}$) and apodization threshold. See \url{https://github.com/justinc97/PGCC_Mask_Generation.git} for more details. 


\section{Forecasts for SO }\label{sec:so_fcast}

As described in section~\ref{subsec:so}, the 
SO LAT survey will cover 16,000\,\sqdeg{} with lower noise levels and improved angular resolution to \planck. 
This low-noise, high-resolution survey will significantly enhance our ability to identify cold clumps and measure their properties in polarization and intensity.

In order to predict the number of polarized clumps that will be detected by SO, we first need to estimate the true number counts on the sky. 
The PGCC sample is a good estimate of the true number counts in the local volume; the  PGCC catalogue is $\sim 70\,$per$\,$cent complete for $S> 300\,$mJy \citep{planck2014-a37}.
Thus we choose to estimate the local luminosity function $n(L)$ from the PGCC catalogue, and scale these numbers to the full Galactic volume observable by SO (with a vertical density profile z(R) described below). 
 Although, this approach relies on the assumption that the number of cold clumps per unit volume is constant, recent results in the literature suggest that the clumps are indeed homogeneously distributed across the Galaxy \citep{alina, planck2014-a37}. 

We estimate the luminosity function, $n(L)$, by counting cold clumps in the PGCC catalogue with measured distances and fluxes above the 70$\,$per$\,$cent completeness threshold. 
The specific criteria are:
\begin{enumerate}
    \item[(i)]  \texttt{DIST\_QUALITY = 1}, i.e., $1\,\sigma $ consistent  distances   measured   with  different methodologies (mainly via NIR and optical extinction);
    \item [(ii)] \texttt{FLUX\_QUALITY = 1}, see Sect. \ref{sec:pgcc}; 
    \item [(iii)] fluxes above the  $70\,$per$\,$cent completeness flux, $300\,$mJy.  
\end{enumerate}	

Once we estimate the luminosity function within the selected dataset, we  apply the correction factor due the incompleteness, by estimating the Galactic volume observable by SO. We assume axial symmetry and  a vertical profile function, $z(R)$, of the Galactocentric radius:
\begin{displaymath}
z(R)  = z_0 \cosh (R/R_0),
\end{displaymath}
with $z_0=100\,$pc, and $R_0= 9\,$kpc  \citep[see][and references therein]{puglisi2017}. The integration  is performed between $R_{min} =0.14 $ and $R_{max}= 19.50\,$kpc. The former  is given by the minimum Galactocentric distance from the selected PGCC entries, whereas the latter is estimated by considering a clump  with size  $8\,$pc \citep[towards the higher end of the observed size distribution, see][]{planck2014-a37} to be resolved  at SO  resolution ($1\,$arcmin).  

\begin{figure}
    \centering
    \includegraphics[width=1\columnwidth]{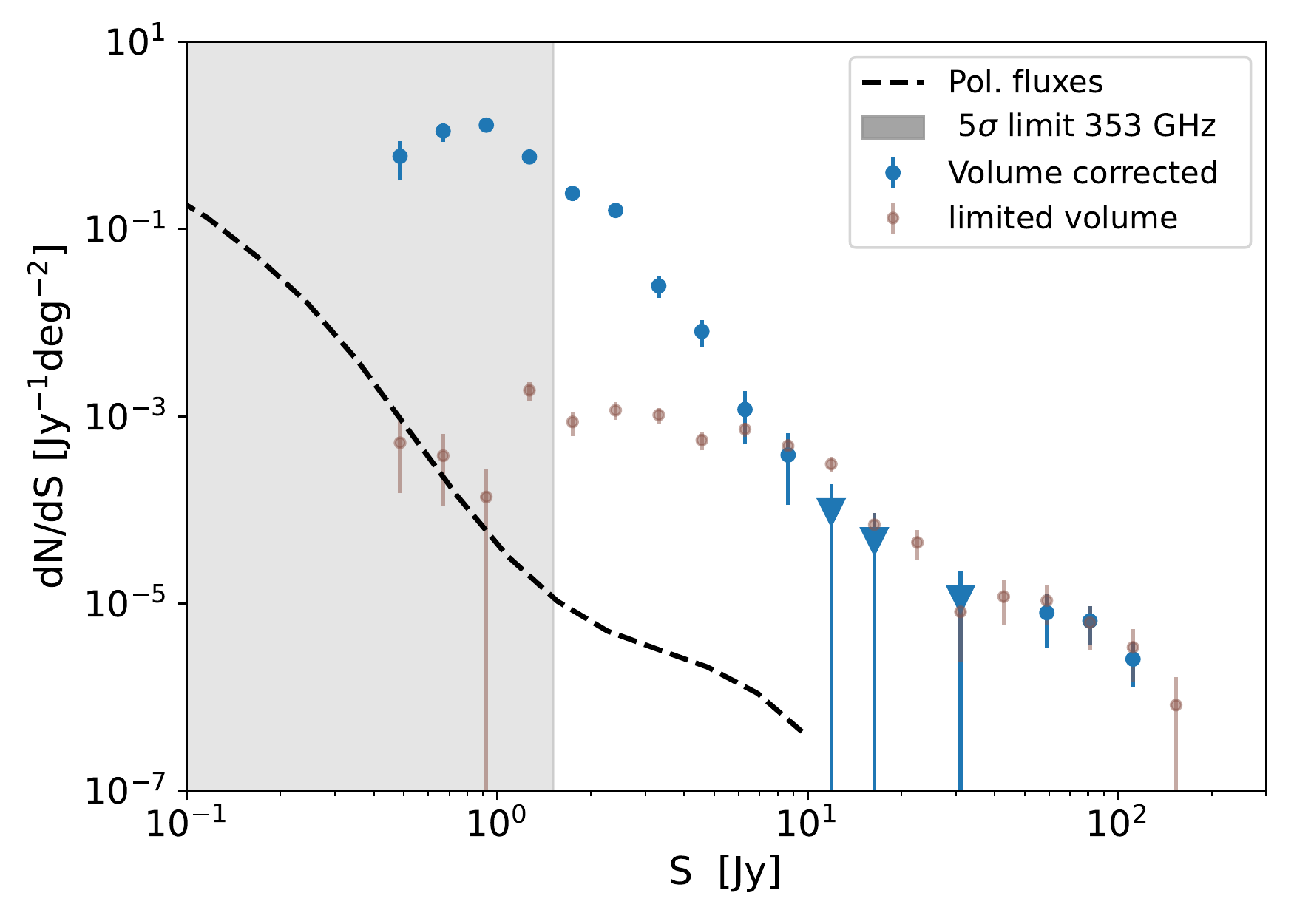}
    
    \caption{Differential number counts of flux densities, $S$,  estimated from the PGCC catalogue at 353$\,$GHz for all the sources satisfying the criteria listed in section~\ref{sec:so_fcast}. The limited volume counts are shown as brown small circles and the counts scaled by Galactic volume are blue large circles. The dashed black line denotes the polarization number counts estimated with equation~\eqref{eq:polcounts}. Upper limits are shown as  (lower triangles). The 70$\,$per$\,$cent completeness of the \textit{Planck} 353$\,$GHz channel is above $\sim1.5\,$Jy so we denote the range below this as the shaded grey area. \label{fig:planck_counts}}
\end{figure}

In Fig.~\ref{fig:planck_counts}, we show the number counts as estimated for both  the limited volume and volume corrected data sets where the error bars are estimated following the prescription of \citet{Gehrels1986}. We note the number counts agree very well at high fluxes $S>10\,$Jy, before diverging as expected to lower fluxes which will tend to be further away. 
We also note the the volume corrected number counts show a power law scaling up to $\sim 100\,$mJy. 

As we are interested in estimating number counts in the SO observing bands centred at 220 and 280\,GHz, rather than the \planck{} 353\,GHz channel, we must convert the 353\,GHz fluxes to the SO frequencies. 
We rescale fluxes assuming the spectral energy distribution (SED) is described by a modified black body function with $\langle \beta_d \rangle = 1.51$ and $\langle T_d\rangle= 10.98$.
These parameters are the average modified black body values for the selected objects in the PGCC catalogue.

We also need to estimate the probability distribution of the polarization fraction, $ \mathcal{P} (\Pi )$, which we assume is  taken from a log-normal distribution. 
 \citet{Puglisi_2018} and \citet{Bonavera2017} showed    $ \mathcal{P} (\Pi )$ can be approximated by a log-normal or a Rice distribution, as these distributions properly  account for the   bias toward higher values due to low sensitivity polarization measurements. 
 We take the measured values reported in section~\ref{sec:results}, i.e. $\Pi^2_{MS}  = [4.79\pm0.44]\times10^{-4}$, as the lognormal parameters. 

\begin{figure}
    \centering
    \includegraphics[width=1\columnwidth]{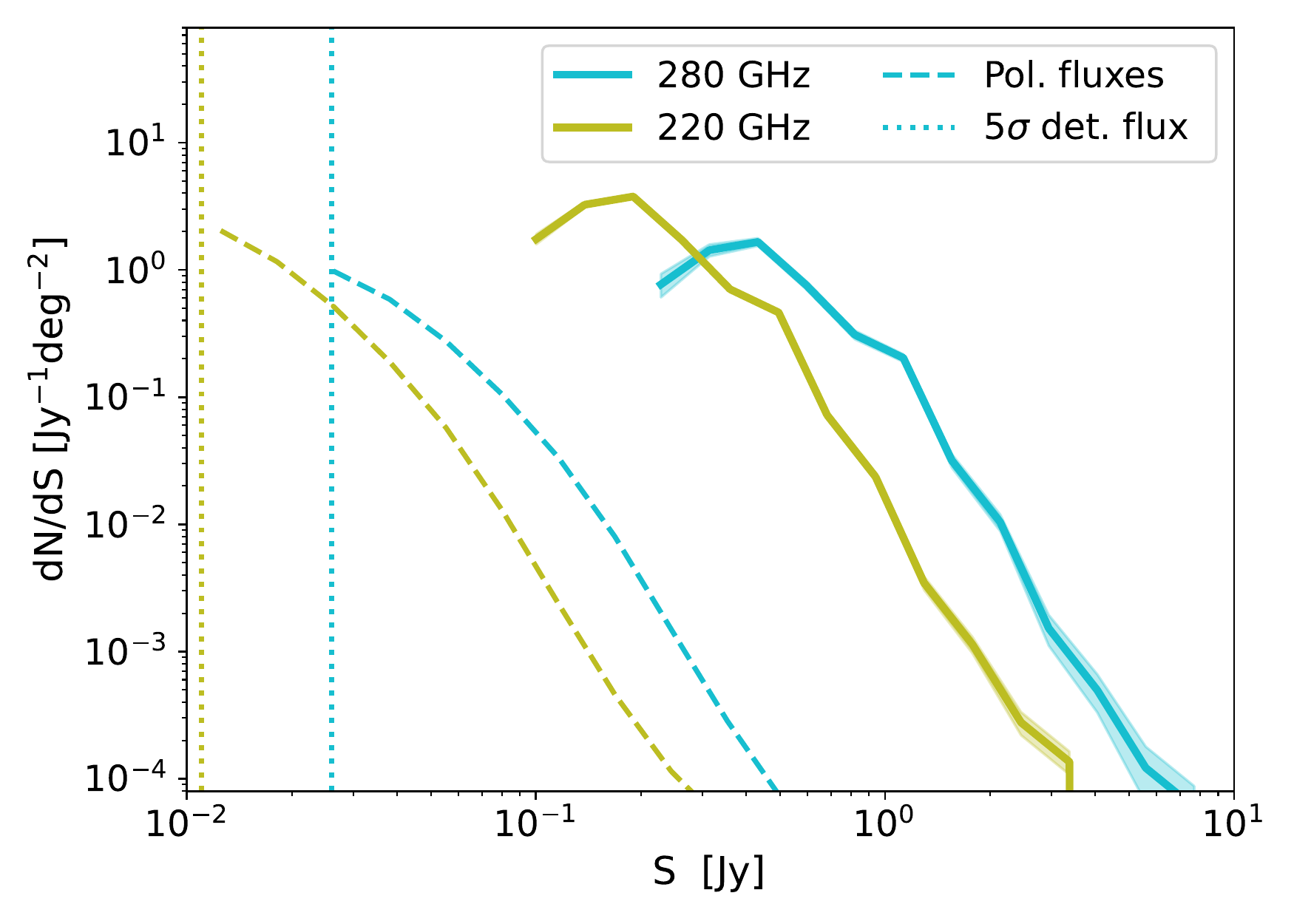}
    
    \caption{Predicted number counts at 220 (green) and 280~GHz (cyan) for the SO survey. The number counts are obtained by rescaling the fluxes at 353$\,$GHz to each frequency using a modified black-body SED. The dashed lines denote the estimated polarization number counts. Error bars are shown as shaded areas. 
    The vertical dashed lines mark the expected $5\,\sigma$ detection limit in polarization in the highest two SO frequency bands. 
    By integrating these curves, we predict that SO will detect approximately 12,000 cold clumps in intensity and $\sim$430 cold clumps in polarization.\label{fig:so_counts}}
\end{figure}

To estimate the polarization number counts, we follow the approach presented in \citet{Puglisi_2018}: 
\begin{equation}
n(P) = N  \int_{S_0=P } ^{\infty} \frac{dS}{S} \mathcal{P} (\Pi, S)=  N  \int_{S_0=P } ^{\infty} \frac{dS}{S} \mathcal{P} (\Pi) n(S).  \label{eq:polcounts}
\end{equation}
Here $N$ indicates the total number of sources having $S >S_0$, $ \mathcal{P} (\Pi, S)$
 is the probability function of finding a source
with flux $S$ and polarized   fraction $\Pi$, and $ \mathcal{P} (\Pi )$ is the probability function of fractional  polarization. 
In the last equality, we have assumed the polarization fraction and flux are statistically independent. 
This is well-justified as we do not observe any correlation between polarization and flux in the current sample (see Fig. \ref{fig:fluxbins} and \ref{fig:luminosity_scatter}).
We insert the volume-corrected number counts and log-normal polarization distribution into  Eqn.~\ref{eq:polcounts} to estimate the number of cold clumps observable by SO. 
We plot the predicted number counts in the SO observing bands in Fig.~\ref{fig:so_counts}.

The SO LAT survey is forecast to have a $5\,\sigma$ detection threshold in polarization of 26 and 11$\,$mJy at 280 and 220$\,$GHz respectively \citep{Ade_2019}. 
At these detection thresholds, we predict SO will detect approximately  
$\sim$430 cold clumps in polarization. 
The expected numbers are very similar between SO's two highest frequency bands, 420 at 220$\,$GHz and 440 at 280$\,$GHz, as both bands have similar sensitivity to objects with a dust-like SED. 
We predict that SO will detect $\sim$12,000 cold clumps in intensity. 
We note that we have not extrapolated the \planck{} number counts to lower fluxes which might increase this number. 
If there are no clumps at lower fluxes than the PGCC catalogue, the SO sensitivity in intensity in both channels is sufficient to detect all cold clumps within the surveyed region of sky.  
 We summarize the number forecasts for the SO survey at both 220 and 280\,GHz in Table~\ref{tab:forecasts}. 
 The predicted number counts as a function of flux are shown in Fig.~\ref{fig:so_counts} with errors in the shaded regions estimated following \citet{Gehrels1986}.
 
  We note that these  forecasts are different with respect to the ones presented in   \citet{Hensley_2022},  as the two forecasts are drawn from two different  catalogues   employing different kind of sources, molecular clouds with size larger than $1\,$pc from \cite{miville_2017}.   
  Reassuringly however,  both methods lead to a similar expected number of detections in intensity at 280\,GHz: 12,000 cold clump detections in this work and 8500 in \citet{Hensley_2022}.

 We also remind the reader  that this forecast relies on the  number counts of polarized sources detected following the criteria listed above and that can be subject to large fluctuations due to the fact that the measured fluxes could be biased toward higher values.
 
 \begin{table}
  
     \centering
  
     \begin{tabular}{cccc}
          \hline
    &   $5\,\sigma_{\rm Pol}$  [mJy] &    $N_{\rm Pol}$     \\
    \hline
220 GHz  & 11 &  420 \\ 
280 GHz  & 26 &  440  \\ 
\hline
     \end{tabular}
     \caption{We predict that SO will detect at $\ge5\,\sigma$ in polarization approximately 430 cold clumps at 220 and 280\,GHz. 
     SO should also detect approximately 12,000 cold clumps in intensity in both bands. 
     The first column lists the expected $5\,\sigma$ polarized flux detection threshold for the SO survey maps at each observing band, while the second column lists the number of sources detected in polarization in each band. 
     \label{tab:forecasts}}
 \end{table}

 The increase of the number of detections from the polarized fluxes of cold clumps will allow us to assess the interplay between the Galactic magnetic field and clumps hosted in filaments. In fact, \citet{alina}   studied the  alignment of filaments with respect to    magnetic fields. They inferred in about 90 locations (selected from the PGCC catalogue) the magnetic field orientations in the filaments hosting a clump and in the surrounding  environment   from the  \emph{Planck} 353$\,$GHz $QU$ maps.  
 
 They observed a trend for low-contrast filaments   to be  parallel to    the background magnetic field;  whereas for  high-contrast filaments  the orientation with respect to the magnetic field is found to be  random.  This might be an indication filaments form and evolve  by strongly interacting with interstellar magnetic fields. 

\begin{figure*} 
     \centering
     \includegraphics[angle=180, width=2\columnwidth]{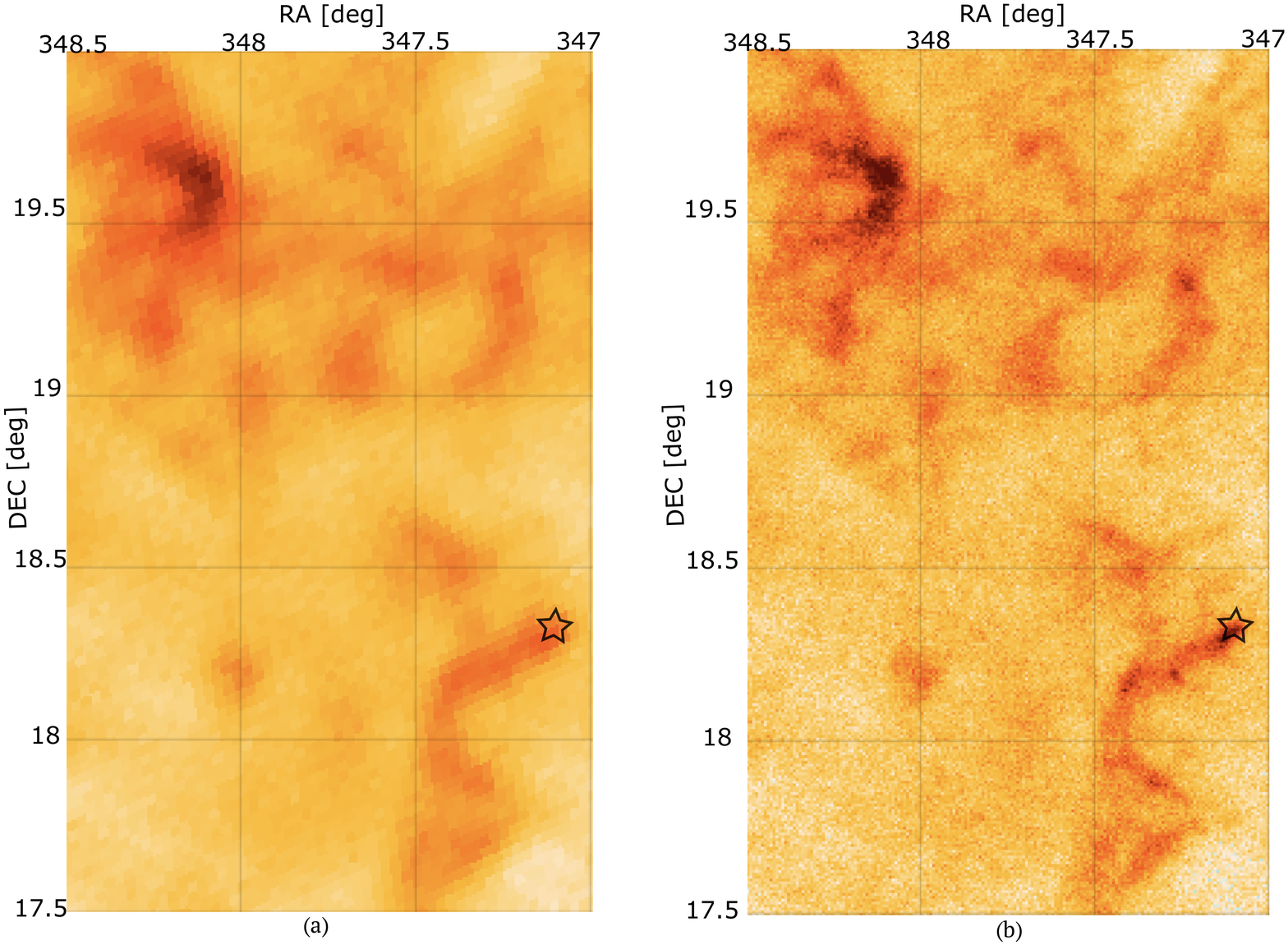}
     \caption{(a) \emph{Planck}  and (b) ACT DR4 temperature maps at 220$\,$GHz centred on the Pegasus complex, hosting the  \texttt{PGCC G91.30-38.16} clump (star). One can easily note two filaments in the top   and bottom right regions of the map.  These filaments are among the 90 detected  by \citet{alina}  using \emph{Planck}  and IRAS 100$\,\mu$m data.}
     \label{fig:planck_act}
 \end{figure*}

To illustrate the value of the SO angular resolution, in Fig. \ref{fig:planck_act}, we show a comparison between the temperature maps of \emph{Planck} and the Atacama Cosmology Telescope \citep[ACT;][]{naess2020} at 220$\,$GHz 
in the vicinity of  a filament hosting the \texttt{PGCC G91.30-38.16} clump. 
The angular resolution of ACT, which is similar to SO's angular resolution and five times smaller than \emph{Planck} at the same frequency, begins to reveal substructures that are washed out by the coarser resolution of \planck. 
The SO LAT survey will cover slightly more sky than ACT (40$\,$per$\,$cent instead of 36$\,$per$\,$cent of the sky), with similar angular resolution ($1\,$arcmin), but significantly lower noise in both temperature and polarization. 
 The low-noise survey from SO will  allow us to carry out a similar analysis as the ones already proposed  in the literature, \citep{alina, cukierman} with unprecedented level of detail in  polarization data. 
 The two SO high frequency channels at 220 and 280$\,$GHz will also allow us to better constrain the spectral properties of filaments and cold clumps.

\section{Conclusion/Summary} \label{sec:conclusion}
We measure for the first time the mean-squared fractional polarization of Galactic cold clumps via stacking in the \textit{Planck} PR4 353$\,$GHz frequency maps. 
We stack the polarization and intensity maps at the positions of reliably detected cold clumps in the PGCC catalogue, and account for background and noise contributions. 
We find the mean-squared polarization fraction of the Galactic cold clumps to be $\langle P^2\rangle/\langle I^2\rangle = [4.79\pm0.44]\times10^{-4}$, giving an 11$\,\sigma$ detection of polarization in Galactic cold clumps. 
We test if the polarization fraction depends on either the source flux or luminosity, and find no evidence that it does. 
We also test if polarization fraction changes with Galactic latitude, where we find the polarization fraction increases at higher Galactic latitudes especially at $|b|>40^\circ$. 
This increase may be due to a physical effect, or alternatively to incomplete background subtraction in one of the subsamples. 
Improved data with more cold clumps at high Galactic latitudes could help distinguish between these explanations. 


Finally, we look at forecasts for the number of cold clumps identifiable by a new experiment, the SO LAT survey of 16,000\,\sqdeg. 
The increased angular resolution and decreased noise compared to the \textit{Planck} satellite will allow for the discovery of cold clumps out to much larger distances and better sample the full distribution of cold clumps across the sky.
We forecast that SO will detect, at $\ge5\,\sigma$ significance, $\sim$12,000 cold clumps in intensity and $\sim$430 cold clumps in polarization. 
As \planck{} detected only $\sim2$ cold clumps at $5\,\sigma$ in polarization, this represents a factor of $\sim$\,200 increase in number counts. 
Future catalogues from SO promise to dramatically improve the statistical weight of all the tests made in this work. 

The lower noise levels and finer angular resolution of the SO survey  will also better resolve any substructure in the cold clumps, in both intensity and polarization. 
\citet{Hensley_2022} previously presented forecasts for the number of similar molecular clouds that SO will resolve in polarization, highlighting this fact.
These resolved studies will allow us to expand our understanding of the magnetic field structure of cold clumps and relationship to their surrounding filaments and molecular clouds. 
Millimeter-wavelength studies of cold clumps may hold the answers to understanding the role of magnetic fields in star forming regions.

\section*{Acknowledgements}
This work was supported in part by a grant from the Simons Foundation (Award \#457687, B.K.). This research used resources of the National Energy Research Scientific Computing Center, which is supported by the Office of Science of the U.S. Department of Energy under Contract No. DE-AC02-05CH11231. 
Melbourne authors acknowledge support from the Australian Research Council's Discovery Projects scheme (DP210102386).
S.E.C. acknowledges support from the National Science Foundation grant No. AST-2106607.
J.C.H. acknowledges support from NSF grant AST-2108536, NASA grant 21-ATP21-0129, DOE grant DE-SC00233966, the Sloan Foundation, and the Simons Foundation. 
The authors would like to thank Juan Diego Soler for useful discussions. 
G. Coppi is supported by the European Research Council under the Marie Skłodowska Curie actions through the Individual European Fellowship No. 892174 PROTOCALC. 
G.F. acknowledges the support of the European Research Council under the Marie Sk\l{}odowska Curie actions through the Individual Global Fellowship No.~892401 PiCOGAMBAS. 
This research has made use of the NASA/IPAC Infrared Science Archive, which is funded by the National Aeronautics and Space Administration and operated by the California Institute of Technology.
Some of the results in this paper have been derived using the healpy and HEALPix packages. 


\section*{Data Availability}
The mask generation code used in this letter is publicly available at \url{https://github.com/justinc97/PGCC_Mask_Generation.git}. We have made masks publicly available with $N_{\textrm{SIDE}}$ resolutions of 2048 and 512, based on the full PGCC catalogue, see the link above for customizable mask generation options. For any bug related reporting, please submit an issue in the GitHub repository.



\bibliographystyle{mnras}
\bibliography{mnras_PGCC} 

\begin{thebibliography}{}
\makeatletter
\relax
\def\mn@urlcharsother{\let\do\@makeother \do\$\do\&\do\#\do\^\do\_\do\%\do\~}
\def\mn@doi{\begingroup\mn@urlcharsother \@ifnextchar [ {\mn@doi@}
  {\mn@doi@[]}}
\def\mn@doi@[#1]#2{\def\@tempa{#1}\ifx\@tempa\@empty \href
  {http://dx.doi.org/#2} {doi:#2}\else \href {http://dx.doi.org/#2} {#1}\fi
  \endgroup}
\def\mn@eprint#1#2{\mn@eprint@#1:#2::\@nil}
\def\mn@eprint@arXiv#1{\href {http://arxiv.org/abs/#1} {{\tt arXiv:#1}}}
\def\mn@eprint@dblp#1{\href {http://dblp.uni-trier.de/rec/bibtex/#1.xml}
  {dblp:#1}}
\def\mn@eprint@#1:#2:#3:#4\@nil{\def\@tempa {#1}\def\@tempb {#2}\def\@tempc
  {#3}\ifx \@tempc \@empty \let \@tempc \@tempb \let \@tempb \@tempa \fi \ifx
  \@tempb \@empty \def\@tempb {arXiv}\fi \@ifundefined
  {mn@eprint@\@tempb}{\@tempb:\@tempc}{\expandafter \expandafter \csname
  mn@eprint@\@tempb\endcsname \expandafter{\@tempc}}}

\bibitem[\protect\citeauthoryear{{Ade} et~al.,}{{Ade} et~al.}{2019}]{Ade_2019}
{Ade} P.,  et~al., 2019, \mn@doi [\jcap] {10.1088/1475-7516/2019/02/056}, \href
  {https://ui.adsabs.harvard.edu/abs/2019JCAP...02..056A} {2019, 056}

\bibitem[\protect\citeauthoryear{{Alina}, {Ristorcelli}, {Montier},
  {Abdikamalov}, {Juvela}, {Ferri{\`e}re}, {Bernard}  \& {Micelotta}}{{Alina}
  et~al.}{2019}]{alina}
{Alina} D.,  {Ristorcelli} I.,  {Montier} L.,  {Abdikamalov} E.,  {Juvela} M.,
  {Ferri{\`e}re} K.,  {Bernard} J.~P.,   {Micelotta} E.~R.,  2019, \mn@doi
  [\mnras] {10.1093/mnras/stz508}, \href
  {https://ui.adsabs.harvard.edu/abs/2019MNRAS.485.2825A} {485, 2825}

\bibitem[\protect\citeauthoryear{{Andr{\'e}} et~al.,}{{Andr{\'e}}
  et~al.}{2010}]{Andre2010}
{Andr{\'e}} P.,  et~al., 2010, \mn@doi [\aap] {10.1051/0004-6361/201014666},
  \href {https://ui.adsabs.harvard.edu/abs/2010A&A...518L.102A} {518, L102}

\bibitem[\protect\citeauthoryear{{Andr{\'e}}, {Di Francesco}, {Ward-Thompson},
  {Inutsuka}, {Pudritz}  \& {Pineda}}{{Andr{\'e}} et~al.}{2014}]{andre2014}
{Andr{\'e}} P.,  {Di Francesco} J.,  {Ward-Thompson} D.,  {Inutsuka} S.~I.,
  {Pudritz} R.~E.,   {Pineda} J.~E.,  2014, in {Beuther} H.,  {Klessen} R.~S.,
  {Dullemond} C.~P.,   {Henning} T.,  eds, Protostars and Planets VI. pp 27--51
  (\mn@eprint {arXiv} {1312.6232}),
  \mn@doi{10.2458/azu_uapress_9780816531240-ch002}

\bibitem[\protect\citeauthoryear{{Bennett} et~al.,}{{Bennett}
  et~al.}{2013}]{Bennett2013}
{Bennett} C.~L.,  et~al., 2013, \mn@doi [\apjs] {10.1088/0067-0049/208/2/20},
  \href {https://ui.adsabs.harvard.edu/abs/2013ApJS..208...20B} {208, 20}

\bibitem[\protect\citeauthoryear{{B{\'e}thermin} et~al.,}{{B{\'e}thermin}
  et~al.}{2012}]{Bethermin2012}
{B{\'e}thermin} M.,  et~al., 2012, \mn@doi [\apjl]
  {10.1088/2041-8205/757/2/L23}, \href
  {https://ui.adsabs.harvard.edu/abs/2012ApJ...757L..23B} {757, L23}

\bibitem[\protect\citeauthoryear{{Bonavera}, {Gonz{\'a}lez-Nuevo},
  {Arg{\"u}eso}  \& {Toffolatti}}{{Bonavera} et~al.}{2017}]{Bonavera2017}
{Bonavera} L.,  {Gonz{\'a}lez-Nuevo} J.,  {Arg{\"u}eso} F.,   {Toffolatti} L.,
  2017, \mn@doi [\mnras] {10.1093/mnras/stx1020}, \href
  {https://ui.adsabs.harvard.edu/abs/2017MNRAS.469.2401B} {469, 2401}

\bibitem[\protect\citeauthoryear{Caselli}{Caselli}{2011}]{caselli2011}
Caselli P.,  2011, \mn@doi [Proceedings of the International Astronomical
  Union] {10.1017/S1743921311024835}, 7, 19–32

\bibitem[\protect\citeauthoryear{{Choi} \& {Page}}{{Choi} \&
  {Page}}{2015}]{Choi2015}
{Choi} S.~K.,  {Page} L.~A.,  2015, \mn@doi [\jcap]
  {10.1088/1475-7516/2015/12/020}, \href
  {https://ui.adsabs.harvard.edu/abs/2015JCAP...12..020C} {2015, 020}

\bibitem[\protect\citeauthoryear{{Cukierman}, {Clark}  \& {Halal}}{{Cukierman}
  et~al.}{2022}]{cukierman}
{Cukierman} A.~J.,  {Clark} S.~E.,   {Halal} G.,  2022, \mn@doi [arXiv
  e-prints] {10.48550/arXiv.2208.07382}, \href
  {https://ui.adsabs.harvard.edu/abs/2022arXiv220807382C} {p. arXiv:2208.07382}

\bibitem[\protect\citeauthoryear{{Dole} et~al.,}{{Dole}
  et~al.}{2006}]{Dole2006}
{Dole} H.,  et~al., 2006, \mn@doi [\aap] {10.1051/0004-6361:20054446}, \href
  {https://ui.adsabs.harvard.edu/abs/2006A&A...451..417D} {451, 417}

\bibitem[\protect\citeauthoryear{{Gehrels}}{{Gehrels}}{1986}]{Gehrels1986}
{Gehrels} N.,  1986, \mn@doi [\apj] {10.1086/164079}, \href
  {https://ui.adsabs.harvard.edu/abs/1986ApJ...303..336G} {303, 336}

\bibitem[\protect\citeauthoryear{{G{\'o}rski}, {Hivon}, {Banday}, {Wandelt},
  {Hansen}, {Reinecke}  \& {Bartelmann}}{{G{\'o}rski}
  et~al.}{2005}]{Gorski2005}
{G{\'o}rski} K.~M.,  {Hivon} E.,  {Banday} A.~J.,  {Wandelt} B.~D.,  {Hansen}
  F.~K.,  {Reinecke} M.,   {Bartelmann} M.,  2005, \mn@doi [\apj]
  {10.1086/427976}, \href
  {https://ui.adsabs.harvard.edu/abs/2005ApJ...622..759G} {622, 759}

\bibitem[\protect\citeauthoryear{{Gudmundsson} et~al.,}{{Gudmundsson}
  et~al.}{2021}]{Gudmundsson_2021}
{Gudmundsson} J.~E.,  et~al., 2021, \mn@doi [\ao] {10.1364/AO.411533}, \href
  {https://ui.adsabs.harvard.edu/abs/2021ApOpt..60..823G} {60, 823}

\bibitem[\protect\citeauthoryear{{Gupta} et~al.,}{{Gupta}
  et~al.}{2019}]{Gupta2019}
{Gupta} N.,  et~al., 2019, \mn@doi [\mnras] {10.1093/mnras/stz2905}, \href
  {https://ui.adsabs.harvard.edu/abs/2019MNRAS.490.5712G} {490, 5712}

\bibitem[\protect\citeauthoryear{{Hacar}, {Clark}, {Heitsch}, {Kainulainen},
  {Panopoulou}, {Seifried}  \& {Smith}}{{Hacar} et~al.}{2022}]{Hacar2022}
{Hacar} A.,  {Clark} S.,  {Heitsch} F.,  {Kainulainen} J.,  {Panopoulou} G.,
  {Seifried} D.,   {Smith} R.,  2022, \mn@doi [arXiv e-prints]
  {10.48550/arXiv.2203.09562}, \href
  {https://ui.adsabs.harvard.edu/abs/2022arXiv220309562H} {p. arXiv:2203.09562}

\bibitem[\protect\citeauthoryear{{Hensley} et~al.,}{{Hensley}
  et~al.}{2022}]{Hensley_2022}
{Hensley} B.~S.,  et~al., 2022, \mn@doi [\apj] {10.3847/1538-4357/ac5e36},
  \href {https://ui.adsabs.harvard.edu/abs/2022ApJ...929..166H} {929, 166}

\bibitem[\protect\citeauthoryear{{Inoue}, {Yamazaki}  \& {Inutsuka}}{{Inoue}
  et~al.}{2009}]{Inoue2009}
{Inoue} T.,  {Yamazaki} R.,   {Inutsuka} S.-i.,  2009, \mn@doi [\apj]
  {10.1088/0004-637X/695/2/825}, \href
  {https://ui.adsabs.harvard.edu/abs/2009ApJ...695..825I} {695, 825}

\bibitem[\protect\citeauthoryear{{Juvela}}{{Juvela}}{2012}]{Juvela2012}
{Juvela} M.,  2012, in From Atoms to Pebbles: Herschel's view of Star and
  Planet Formation. p.~16

\bibitem[\protect\citeauthoryear{{Juvela} et~al.,}{{Juvela}
  et~al.}{2015}]{Juvela2015}
{Juvela} M.,  et~al., 2015, \mn@doi [\aap] {10.1051/0004-6361/201423788}, \href
  {https://ui.adsabs.harvard.edu/abs/2015A&A...584A..93J} {584, A93}

\bibitem[\protect\citeauthoryear{{Juvela} et~al.,}{{Juvela}
  et~al.}{2018a}]{Juvela2018b}
{Juvela} M.,  et~al., 2018a, \mn@doi [\aap] {10.1051/0004-6361/201731921},
  \href {https://ui.adsabs.harvard.edu/abs/2018A&A...612A..71J} {612, A71}

\bibitem[\protect\citeauthoryear{{Juvela}, {Malinen}, {Montillaud}, {Pelkonen},
  {Ristorcelli}  \& {T{\'o}th}}{{Juvela} et~al.}{2018b}]{Juvela2018a}
{Juvela} M.,  {Malinen} J.,  {Montillaud} J.,  {Pelkonen} V.~M.,  {Ristorcelli}
  I.,   {T{\'o}th} L.~V.,  2018b, \mn@doi [\aap] {10.1051/0004-6361/201630304},
  \href {https://ui.adsabs.harvard.edu/abs/2018A&A...614A..83J} {614, A83}

\bibitem[\protect\citeauthoryear{{Kamionkowski}, {Kosowsky}  \&
  {Stebbins}}{{Kamionkowski} et~al.}{1997}]{Kamionkowski1996}
{Kamionkowski} M.,  {Kosowsky} A.,   {Stebbins} A.,  1997, \mn@doi [\prl]
  {10.1103/PhysRevLett.78.2058}, \href
  {https://ui.adsabs.harvard.edu/abs/1997PhRvL..78.2058K} {78, 2058}

\bibitem[\protect\citeauthoryear{{Liu} et~al.,}{{Liu} et~al.}{2018}]{Liu2018}
{Liu} T.,  et~al., 2018, \mn@doi [\apjs] {10.3847/1538-4365/aaa3dd}, \href
  {https://ui.adsabs.harvard.edu/abs/2018ApJS..234...28L} {234, 28}

\bibitem[\protect\citeauthoryear{{Miville-Desch{\^e}nes} \&
  {Lagache}}{{Miville-Desch{\^e}nes} \&
  {Lagache}}{2005}]{Miville-Deschenes2004}
{Miville-Desch{\^e}nes} M.-A.,  {Lagache} G.,  2005, \mn@doi [\apjs]
  {10.1086/427938}, \href
  {https://ui.adsabs.harvard.edu/abs/2005ApJS..157..302M} {157, 302}

\bibitem[\protect\citeauthoryear{{Miville-Desch{\^e}nes}, {Murray}  \&
  {Lee}}{{Miville-Desch{\^e}nes} et~al.}{2017}]{miville_2017}
{Miville-Desch{\^e}nes} M.-A.,  {Murray} N.,   {Lee} E.~J.,  2017, \mn@doi
  [\apj] {10.3847/1538-4357/834/1/57}, \href
  {https://ui.adsabs.harvard.edu/abs/2017ApJ...834...57M} {834, 57}

\bibitem[\protect\citeauthoryear{{Montier} \& {Giard}}{{Montier} \&
  {Giard}}{2005}]{Montier2005}
{Montier} L.~A.,  {Giard} M.,  2005, \mn@doi [\aap]
  {10.1051/0004-6361:20042388}, \href
  {https://ui.adsabs.harvard.edu/abs/2005A&A...439...35M} {439, 35}

\bibitem[\protect\citeauthoryear{{Montillaud} et~al.,}{{Montillaud}
  et~al.}{2015}]{Montillaud2015}
{Montillaud} J.,  et~al., 2015, \mn@doi [\aap] {10.1051/0004-6361/201424063},
  \href {https://ui.adsabs.harvard.edu/abs/2015A&A...584A..92M} {584, A92}

\bibitem[\protect\citeauthoryear{{Myers} \& {Goodman}}{{Myers} \&
  {Goodman}}{1991}]{Myers1991}
{Myers} P.~C.,  {Goodman} A.~A.,  1991, \mn@doi [\apj] {10.1086/170070}, \href
  {https://ui.adsabs.harvard.edu/abs/1991ApJ...373..509M} {373, 509}

\bibitem[\protect\citeauthoryear{{Naess} et~al.,}{{Naess}
  et~al.}{2020}]{naess2020}
{Naess} S.,  et~al., 2020, \mn@doi [\jcap] {10.1088/1475-7516/2020/12/046},
  \href {https://ui.adsabs.harvard.edu/abs/2020JCAP...12..046N} {2020, 046}

\bibitem[\protect\citeauthoryear{{Nakano} \& {Nakamura}}{{Nakano} \&
  {Nakamura}}{1978}]{NakanoandNakamura1978}
{Nakano} T.,  {Nakamura} T.,  1978, \pasj, \href
  {https://ui.adsabs.harvard.edu/abs/1978PASJ...30..671N} {30, 671}

\bibitem[\protect\citeauthoryear{{Neugebauer} et~al.,}{{Neugebauer}
  et~al.}{1984}]{Neugebauer1984}
{Neugebauer} G.,  et~al., 1984, \mn@doi [\apjl] {10.1086/184209}, \href
  {https://ui.adsabs.harvard.edu/abs/1984ApJ...278L...1N} {278, L1}

\bibitem[\protect\citeauthoryear{{Page} et~al.,}{{Page}
  et~al.}{2007}]{Page2007}
{Page} L.,  et~al., 2007, \mn@doi [\apjs] {10.1086/513699}, \href
  {https://ui.adsabs.harvard.edu/abs/2007ApJS..170..335P} {170, 335}

\bibitem[\protect\citeauthoryear{{\sorthelp{Planck Collaboration 2015J}}{Planck
  Collaboration X}}{{\sorthelp{Planck Collaboration 2015J}}{Planck
  Collaboration X}}{2016}]{planck2014-a12}
{\sorthelp{Planck Collaboration 2015J}}{Planck Collaboration X} 2016, \mn@doi
  [\aap] {10.1051/0004-6361/201525967}, \href
  {https://ui.adsabs.harvard.edu/abs/2016A&A...594A..10P} {594, A10}

\bibitem[\protect\citeauthoryear{{\sorthelp{Planck Collaboration
  2015ZC}}{Planck Collaboration XXVIII}}{{\sorthelp{Planck Collaboration
  2015ZC}}{Planck Collaboration XXVIII}}{2016}]{planck2014-a37}
{\sorthelp{Planck Collaboration 2015ZC}}{Planck Collaboration XXVIII} 2016,
  \mn@doi [\aap] {10.1051/0004-6361/201525819}, \href
  {https://ui.adsabs.harvard.edu/abs/2016A&A...594A..28P} {594, A28}

\bibitem[\protect\citeauthoryear{{\sorthelp{Planck Collaboration 2018A}}{Planck
  Collaboration I}}{{\sorthelp{Planck Collaboration 2018A}}{Planck
  Collaboration I}}{2020}]{planck2016-l01}
{\sorthelp{Planck Collaboration 2018A}}{Planck Collaboration I} 2020, \mn@doi
  [\aap] {10.1051/0004-6361/201833880}, \href
  {https://ui.adsabs.harvard.edu/abs/2020A&A...641A...1P} {641, A1}

\bibitem[\protect\citeauthoryear{{\sorthelp{Planck Collaboration 2018C}}{Planck
  Collaboration III}}{{\sorthelp{Planck Collaboration 2018C}}{Planck
  Collaboration III}}{2020}]{planck2016-l03}
{\sorthelp{Planck Collaboration 2018C}}{Planck Collaboration III} 2020, \mn@doi
  [\aap] {10.1051/0004-6361/201832909}, \href
  {https://ui.adsabs.harvard.edu/abs/2020A&A...641A...3P} {641, A3}

\bibitem[\protect\citeauthoryear{{\sorthelp{Planck Collaboration 2018D}}{Planck
  Collaboration IV}}{{\sorthelp{Planck Collaboration 2018D}}{Planck
  Collaboration IV}}{2020}]{planck2016-l04}
{\sorthelp{Planck Collaboration 2018D}}{Planck Collaboration IV} 2020, \mn@doi
  [\aap] {10.1051/0004-6361/201833881}, \href
  {https://ui.adsabs.harvard.edu/abs/2020A&A...641A...4P} {641, A4}

\bibitem[\protect\citeauthoryear{{\sorthelp{Planck Collaboration 2018K}}{Planck
  Collaboration XI}}{{\sorthelp{Planck Collaboration 2018K}}{Planck
  Collaboration XI}}{2020}]{planck2016-l11A}
{\sorthelp{Planck Collaboration 2018K}}{Planck Collaboration XI} 2020, \mn@doi
  [\aap] {10.1051/0004-6361/201832618}, \href
  {https://ui.adsabs.harvard.edu/abs/2020A&A...641A..11P} {641, A11}

\bibitem[\protect\citeauthoryear{{\sorthelp{Planck Collaboration 2018L}}{Planck
  Collaboration XII}}{{\sorthelp{Planck Collaboration 2018L}}{Planck
  Collaboration XII}}{2020}]{planck2016-l11B}
{\sorthelp{Planck Collaboration 2018L}}{Planck Collaboration XII} 2020, \mn@doi
  [\aap] {10.1051/0004-6361/201833885}, \href
  {https://ui.adsabs.harvard.edu/abs/2020A&A...641A..12P} {641, A12}

\bibitem[\protect\citeauthoryear{{\sorthelp{Planck Collaboration IntZM}}{Planck
  Collaboration Int. XXXVIII}}{{\sorthelp{Planck Collaboration IntZM}}{Planck
  Collaboration Int. XXXVIII}}{2016}]{planck2015-XXXVIII}
{\sorthelp{Planck Collaboration IntZM}}{Planck Collaboration Int. XXXVIII}
  2016, \mn@doi [\aap] {10.1051/0004-6361/201526506}, \href
  {https://ui.adsabs.harvard.edu/abs/2016A&A...586A.141P} {586, A141}

\bibitem[\protect\citeauthoryear{{\sorthelp{Planck Collaboration IntZW}}{Planck
  Collaboration Int. XLVIII}}{{\sorthelp{Planck Collaboration IntZW}}{Planck
  Collaboration Int. XLVIII}}{2016}]{planck2016-XLVIII}
{\sorthelp{Planck Collaboration IntZW}}{Planck Collaboration Int. XLVIII} 2016,
  \mn@doi [\aap] {10.1051/0004-6361/201629022}, \href
  {https://ui.adsabs.harvard.edu/abs/2016A&A...596A.109P} {596, A109}

\bibitem[\protect\citeauthoryear{{\sorthelp{Planck Collaboration
  IntZZG}}{Planck Collaboration Int. LVII}}{{\sorthelp{Planck Collaboration
  IntZZG}}{Planck Collaboration Int. LVII}}{2020}]{planck2020-LVII}
{\sorthelp{Planck Collaboration IntZZG}}{Planck Collaboration Int. LVII} 2020,
  \mn@doi [\aap] {10.1051/0004-6361/202038073}, \href
  {https://ui.adsabs.harvard.edu/abs/2020A&A...643A..42P} {643, 42}

\bibitem[\protect\citeauthoryear{{Puglisi}, {Fabbian}  \&
  {Baccigalupi}}{{Puglisi} et~al.}{2017}]{puglisi2017}
{Puglisi} G.,  {Fabbian} G.,   {Baccigalupi} C.,  2017, \mn@doi [\mnras]
  {10.1093/mnras/stx1029}, \href
  {https://ui.adsabs.harvard.edu/abs/2017MNRAS.469.2982P} {469, 2982}

\bibitem[\protect\citeauthoryear{{Puglisi} et~al.,}{{Puglisi}
  et~al.}{2018}]{Puglisi_2018}
{Puglisi} G.,  et~al., 2018, \mn@doi [\apj] {10.3847/1538-4357/aab3c7}, \href
  {https://ui.adsabs.harvard.edu/abs/2018ApJ...858...85P} {858, 85}

\bibitem[\protect\citeauthoryear{{Rivera-Ingraham} et~al.,}{{Rivera-Ingraham}
  et~al.}{2016}]{Rivera-Ingraham2016}
{Rivera-Ingraham} A.,  et~al., 2016, \mn@doi [\aap]
  {10.1051/0004-6361/201526263}, \href
  {https://ui.adsabs.harvard.edu/abs/2016A&A...591A..90R} {591, A90}

\bibitem[\protect\citeauthoryear{{Seifried} \& {Walch}}{{Seifried} \&
  {Walch}}{2015}]{SeifriedandWalch2015}
{Seifried} D.,  {Walch} S.,  2015, \mn@doi [\mnras] {10.1093/mnras/stv1458},
  \href {https://ui.adsabs.harvard.edu/abs/2015MNRAS.452.2410S} {452, 2410}

\bibitem[\protect\citeauthoryear{{Seljak} \& {Zaldarriaga}}{{Seljak} \&
  {Zaldarriaga}}{1997}]{Seljak1996}
{Seljak} U.,  {Zaldarriaga} M.,  1997, \mn@doi [\prl]
  {10.1103/PhysRevLett.78.2054}, \href
  {https://ui.adsabs.harvard.edu/abs/1997PhRvL..78.2054S} {78, 2054}

\bibitem[\protect\citeauthoryear{{Trombetti}, {Burigana}, {De Zotti},
  {Galluzzi}  \& {Massardi}}{{Trombetti} et~al.}{2018}]{Trombetti2018}
{Trombetti} T.,  {Burigana} C.,  {De Zotti} G.,  {Galluzzi} V.,   {Massardi}
  M.,  2018, \mn@doi [\aap] {10.1051/0004-6361/201732342}, \href
  {https://ui.adsabs.harvard.edu/abs/2018A&A...618A..29T} {618, A29}

\bibitem[\protect\citeauthoryear{{Wakelam}, {Gratier}, {Ruaud}, {Le Gal},
  {Majumdar}, {Loison}  \& {Hickson}}{{Wakelam} et~al.}{2021}]{Wakelam2021}
{Wakelam} V.,  {Gratier} P.,  {Ruaud} M.,  {Le Gal} R.,  {Majumdar} L.,
  {Loison} J.~C.,   {Hickson} K.~M.,  2021, \mn@doi [\aap]
  {10.1051/0004-6361/202039367}, \href
  {https://ui.adsabs.harvard.edu/abs/2021A&A...647A.172W} {647, A172}

\bibitem[\protect\citeauthoryear{{Ward-Thompson}, {Scott}, {Hills}  \&
  {Andre}}{{Ward-Thompson} et~al.}{1994}]{ward-thompson1994}
{Ward-Thompson} D.,  {Scott} P.~F.,  {Hills} R.~E.,   {Andre} P.,  1994,
  \mn@doi [\mnras] {10.1093/mnras/268.1.276}, \href
  {https://ui.adsabs.harvard.edu/abs/1994MNRAS.268..276W} {268, 276}

\bibitem[\protect\citeauthoryear{{Xu} et~al.,}{{Xu} et~al.}{2021}]{Xu2021}
{Xu} Z.,  et~al., 2021, \mn@doi [Research Notes of the American Astronomical
  Society] {10.3847/2515-5172/abf9ab}, \href
  {https://ui.adsabs.harvard.edu/abs/2021RNAAS...5..100X} {5, 100}

\bibitem[\protect\citeauthoryear{{Zahorecz}, {Jimenez-Serra}, {Wang}, {Testi},
  {T{\'o}th}  \& {Molinari}}{{Zahorecz} et~al.}{2016}]{Zahorecz2016}
{Zahorecz} S.,  {Jimenez-Serra} I.,  {Wang} K.,  {Testi} L.,  {T{\'o}th} L.~V.,
    {Molinari} S.,  2016, \mn@doi [\aap] {10.1051/0004-6361/201527909}, \href
  {https://ui.adsabs.harvard.edu/abs/2016A&A...591A.105Z} {591, A105}

\bibitem[\protect\citeauthoryear{{Zonca}, {Singer}, {Lenz}, {Reinecke},
  {Rosset}, {Hivon}  \& {Gorski}}{{Zonca} et~al.}{2019}]{Zonca2019}
{Zonca} A.,  {Singer} L.,  {Lenz} D.,  {Reinecke} M.,  {Rosset} C.,  {Hivon}
  E.,   {Gorski} K.,  2019, \mn@doi [The Journal of Open Source Software]
  {10.21105/joss.01298}, \href
  {https://ui.adsabs.harvard.edu/abs/2019JOSS....4.1298Z} {4, 1298}

\makeatother
\end{thebibliography}
\appendix









\bsp	
\label{lastpage}
\end{document}